\documentclass[iop]{emulateapj}

\usepackage{natbib,aas_macros,amsmath}
\usepackage{natbib}

\usepackage{multirow,color}

\newcommand{\cii}{[C\,{\sc ii}]}

\begin{document}
\shorttitle{
Systematic Measurements of \cii\ Size \& Morphology at $z=4-6$
}
\shortauthors{Fujimoto et al.}
\slugcomment{ApJ in press}

\title{%
The ALPINE -- ALMA \cii\ Survey: Size of individual star-forming galaxies \\ at $\lowercase{z}=4-6$ 
and their extended halo structure
}

\author{%
Seiji Fujimoto\altaffilmark{1,2,3,4},
John D. Silverman\altaffilmark{5,6},
Matthieu Bethermin\altaffilmark{7}, 
Michele Ginolfi\altaffilmark{8},
Gareth C. Jones\altaffilmark{9,10}, 
Olivier Le F\`{e}vre\altaffilmark{7}
Miroslava Dessauges-Zavadsky\altaffilmark{8}, 
Wiphu Rujopakarn\altaffilmark{11,12}, 
Andreas L. Faisst\altaffilmark{13},
Yoshinobu Fudamoto\altaffilmark{8}, 
Paolo Cassata\altaffilmark{14,15}, 
Laura Morselli\altaffilmark{14,15},
Roberto Maiolino\altaffilmark{9,10},  
Daniel Schaerer\altaffilmark{8}, 
Peter Capak\altaffilmark{16}, 
Lin Yan\altaffilmark{16}, 
Livia Vallini\altaffilmark{17}, 
Sune Toft\altaffilmark{1,2}, 
Federica Loiacono\altaffilmark{18,19}, 
Gianni Zamorani\altaffilmark{19}, 
Margherita Talia\altaffilmark{18,19}, 
Desika Narayanan \altaffilmark{20,1,2}, 
Nimish P. Hathi\altaffilmark{21},
Brian C. Lemaux\altaffilmark{22}, 
M\'{e}d\'{e}ric Boquien\altaffilmark{23}, 
Ricardo Amorin\altaffilmark{24,25}, 
Edo Ibar\altaffilmark{26}, 
Anton M. Koekemoer\altaffilmark{21},
Hugo M\'{e}ndez-Hern\'{a}ndez\altaffilmark{26}, 
Sandro Bardelli\altaffilmark{19}, 
Daniela Vergani\altaffilmark{19}, 
Elena Zucca\altaffilmark{19}, 
Michael Romano\altaffilmark{14,15}, and 
Andrea Cimatti\altaffilmark{18,27}
}

\email{fujimoto@nbi.ku.dk}

\affil{$^{1}$
Cosmic Dawn Center (DAWN), Copenhagen, Denmark
}

\affil{$^{2}$
Niels Bohr Institute, University of Copenhagen, Lyngbyvej 2, DK2100 Copenhagen, Denmark
}

\affil{$^{3}$
Research Institute for Science and Engineering, Waseda University, 3-4-1 Okubo, Shinjuku, Tokyo 169-8555, Japan
}

\affil{$^{4}$
National Astronomical Observatory of Japan, 2-21-1, Osawa, Mitaka, Tokyo, Japan
}

\affil{$^{5}$
Kavli Institute for the Physics and Mathematics of the Universe,
The University of Tokyo, Kashiwa, Japan 277-8583
}

\affil{$^{6}$
Department of Astronomy, School of Science, The University
of Tokyo, 7-3-1 Hongo, Bunkyo, Tokyo 113-0033, Japan
}

\affil{$^{7}$
Aix Marseille Univ, CNRS, CNES, LAM, Marseille, France
}

\affil{$^{8}$
Observatoire de Gen\`eve, Universit\'{e} de Gen\`{e}ve, 51 Ch. des Maillettes, 1290 Versoix, Switzerland
}

\affil{$^{9}$
Cavendish Laboratory, University of Cambridge, 19 J. J. Thomson Ave., Cambridge CB3 0HE, UK
}

\affil{$^{10}$
Kavli Institute for Cosmology, University of Cambridge, Madingley Road, Cambridge CB3 0HA, UK
}

\affil{$^{11}$
Department of Physics, Faculty of Science, Chulalongkorn University, 254 Phayathai Road, Pathumwan, Bangkok 10330, Thailand
}

\affil{$^{12}$
National Astronomical Research Institute of Thailand (Public Organization), Don Kaeo, Mae Rim, Chiang Mai 50180, Thailand
}

\affil{$^{13}$
IPAC, M/C 314-6, California Institute of Technology, 1200 East California Boulevard, Pasadena, CA 91125, USA
}

\affil{$^{14}$
Dipartimento di Fisica e Astronomia, Universit\`{a} di Padova,
Vicolo dell' Osservatorio, 3 35122 Padova, Italy
}

\affil{$^{15}$
INAF Osservatorio Astronomico di Padova, Vicolo
dell' Osservatorio 5, I-35122 Padova, Italy
}

\affil{$^{16}$
The Caltech Optical Observatories, California Institute of Technology, Pasadena, CA 91125, USA
}

\affil{$^{17}$
Leiden Observatory, Leiden University, PO Box 9500, 2300 RA Leiden, The Netherlands.
}

\affil{$^{18}$
University of Bologna, Department of Physics and Astronomy (DIFA), Via Gobetti 93/2, I-40129, Bologna, Italy
}

\affil{$^{19}$
Osservatorio di Astrofisica e Scienza dello Spazio - Istituto Nazionale di Astrofisica, via Gobetti 93/3, I-40129, Bologna, Italy
}

\affil{$^{20}$
Department of Astronomy, University of Florida, 211 Bryant Space Sciences Center, Gainesville, FL 32607
}

\affil{$^{21}$
Space Telescope Science Institute, 3700 San Martin Drive, Baltimore 21218, USA
}

\affil{$^{22}$
Department of Physics, University of California, Davis, One Shields Ave., Davis, CA 95616, USA
}

\affil{$^{23}$
Centro de Astronom\'{i}a (CITEVA), Universidad de Antofagasta, Avenida Angamos 601, Antofagasta, Chile
}

\affil{$^{24}$
Departamento de Astronomia, Universidad de La Serena, Av. Juan Cisternas 1200 Norte, La Serena, Chile
}

\affil{$^{25}$
Instituto de Investigaci\'{o}n Multidisciplinar en Ciencia y Tecnolog\'{i}a, Universidad de La Serena, Ra\'{u}l Bitr\'{a}n 1305, La Serena, Chile
}

\affil{$^{26}$
Instituto de F\'{i}sica y Astronom\'{i}a, Universidad de Valpara\'{i}so, Avda. Gran Breta\~na 1111, Valpara\'{i}0so, Chile
}

\affil{$^{27}$
INAF - Osservatorio Astrofisico di Arcetri, Largo E. Fermi 5, I-50125, Firenze, Italy
}

\def\apj{ApJ}%
\def\apjl{ApJ}%
\def\apjs{ApJS}%

\def\rme{\rm e}
\def\rmstar{\rm star}
\def\rmFIR{\rm FIR}
\def\itHubble{\it Hubble}
\def\rmyr{\rm yr}

\begin{abstract}
We present the physical extent of \cii\ 158$\mu$m line-emitting gas from 46 star-forming galaxies at $z=4-6$ from the ALMA Large Program to INvestigate CII at Early Times (ALPINE). 
Using exponential profile fits, we measure the effective radius of the \cii\ line ($r_{\rme,[CII]}$) for individual galaxies and compare them with the rest-frame ultra-violet (UV) continuum ($r_{\rme,UV}$) from {\it\,Hubble\,\,Space\,\,Telescope} images.  
The effective radius $r_{\rme,[CII]}$ exceeds $r_{\rme,UV}$ by factors of $\sim2$--3 and the ratio of $r_{\rme,[CII]}/r_{\rme,UV}$ increases as a function of $M_{\rmstar}$.  
We do not find strong evidence that \cii\ line, the rest-frame UV, and FIR continuum are always displaced over $\simeq$ 1-kpc scale from each other. 
We identify 30\% of isolated ALPINE sources as having an extended \cii\ component over 10-kpc scales detected at 4.1$\sigma$--10.9$\sigma$ beyond the size of rest-frame UV and far-infrared (FIR) continuum.  
One object has tentative rotating features up to $\sim10$ kpc, where the 3D model fit shows the rotating \cii-gas disk spread over 4 times larger than the rest-frame UV-emitting region. 
Galaxies with the extended \cii\ line structure have high star-formation rate (SFR), stellar mass ($M_{\rmstar}$), low Ly$\alpha$ equivalent-width, and more blue-shifted (red-shifted) rest-frame UV metal absorption (Ly$\alpha$ line), as compared to galaxies without such extended \cii\ structures. 
Although we cannot rule out the possibility that a selection bias towards luminous objects may be responsible for such trends, 
the star-formation driven outflow also explains all these trends.
Deeper observations are essential to test whether the extended \cii\ line structures are ubiquitous to high-$z$ star-forming galaxies.
\end{abstract}

\section{Introduction}
\label{sec:intro}

The first-generation of galaxies form in high density peaks of primordial gas composed of light elements such as hydrogen and helium. Heavy elements are then produced in these galaxies from stellar nucleosynthesis, and are ejected by feedback processes such as outflows driven by the thermal heating and kinetic feedback of supernova explosions \citep[e.g.,][]{pallottini2014, turner2017}.  
Since galaxies evolve within an overlying circum-galactic medium (CGM) impacted by frequent mergers, inflow, and outflow of gas, 
detailed study of galaxy sizes and morphologies at early epochs provides invaluable information on the galaxy assembly in such a crucial phase of galaxy evolution.

On the CGM scale ( $\sim1$--10 kpc) at early cosmic times, 
it has been revealed that star-forming galaxies are surrounded by extended Ly$\alpha$ line emission, namely the Ly$\alpha$ halo, that is on average 10 times larger than their corresponding galaxy sizes in the rest-frame ultra-violet (UV) wavelength \citep[e.g.,][]
{momose2016,wisotzki2016, leclercq2017}.  
While Ly$\alpha$ is fundamental to probe the neutral hydrogen distribution around galaxies, 
only metal lines allow us to probe the enriched gas distribution and hence constrain the efficiency of feedback and star formation out to CGM scales.
Although the rest-frame UV spectrum offers the possibility to detect metal nebular lines such as C{\sc iv} and C{\sc iii]}, these metal lines are so weak that they can be detected in the CGM scale only when a quasar contributes to the ionization \citep[e.g.,][]{guo2020}. 

The metal gas emission and the rest-frame far-infrared (FIR) dust continuum can now be probed up to $z\sim9$ efficiently with the Atacama Large Millimeter/submillimeter Array (ALMA)  \citep[e.g.,][]{watson2015,maiolino2015,capak2015,pentericci2016,knudsen2016,matthee2017,carniani2018b,smit2018,bowler2018,hashimoto2018,hashimoto2019,tamura2019,fujimoto2019}. 
On scales of the CGM, Fujimoto et al. (2019) reported the existence of extended \cii\ line structure over a radius of 10 kpc via the stacking analysis of 18 star-forming galaxies at $z=5.152-7.142$.
The radial surface brightness profile exhibits a 10-kpc scale {\sc [Cii]} halo at the 9.2$\sigma$ level, significantly more extended than the {\it Hubble \, Space \, Telescope} (HST) stellar continuum emission and the dust continuum. 
Interestingly, the radial profiles of the {\sc [Cii]} and Ly$\alpha$ halos show a good agreement with each other,
suggesting a possible link between these halos. 
However, the physical origin of the \cii\ halo, potentially associated with the Ly$\alpha$ halo, is still an open question.
With the {\it ALMA Large Program to Investigate C$^{+}$ at Early Times} (ALPINE), 
the extended \cii\ halo has also been detected in an independent stacking approach \citep{ginolfi2020}. 
Even so, individual observations are essential to further understand the physical origins of the metal gas distributions. 

In this paper, we study the individual size and structure of {\sc [Cii]} emission from star-forming galaxies at $z=4-6$,
drawn from the ALPINE survey. Making full use of the large ancillary dataset including deep HST images, we examine the distributions of the star-forming regions and the metal-enriched gas up to the CGM scale in the early Universe.  
The organization of this paper is as follows. 
In Section 2, the overview and the data products of the ALPINE survey are described.  
Section 3 outlines the method of the size measurements for the ALMA and HST data.  
We report the results of the size measurements and associated physical properties in Section 4. 
In Section 5, we discuss the physical origin of the extended {\sc [Cii]} line emission.  
A summary of this study is presented in Section 6. 
Throughout this paper, we assume a flat universe with 
$\Omega_{\rm m} = 0.3$, 
$\Omega_\Lambda = 0.7$, 
$\sigma_8 = 0.8$, 
and $H_0 = 70$ km s$^{-1}$ Mpc$^{-1}$. 

\section{ALPINE Survey} 
\label{sec:alpine}

\subsection{Overview}
\label{sec:overview}

ALPINE is an ALMA large program (Project ID: 2017.1.00428.L; PI: O. Le F\`evre; see also : e.g., \citealt{lefevre2019,faisst2020,bethermin2020}), 
aimed at measuring \cii\ 158 $\mu$m and rest-frame FIR continuum emission from a representative sample of 118 main-sequence galaxies at $4.4<z<5.8$. 
Observations were carried out between May 2018 and February 2019. 
All galaxies are located in the regions of the Cosmic Evolution Survey (COSMOS; \citealt{scoville2007,koekemoer2007}), The Great Observations Origins Deep Survey South (GOODS-S; \citealt{giavalisco2004}), and CANDELS (Grogin et al. 2011; Koekemoer et al 2011). 
They have star-formation rates (SFR) $\gtrsim$ 10 $M_{\odot}\,{\rm yr^{-1}}$ and stellar masses $M_{\rm star}\sim10^{9}-10^{11}\,M_{\odot}$, generally falling on the main sequence \citep[e.g.,][]{speagle2014,steinhardt2014}. 
The \cii\ line frequency was covered in one tuning based on the spectroscopic redshift as estimated from previous rest-frame UV spectroscopy 
including the VUDS and DEIMOS surveys \citep[e.g.,][]{lefevre2015}.
SFR and $M_{\rm star}$ are derived from template SED fitting using the ancillary photometry available data sets from the rest-frame UV to IR wavelengths. 
The data sets include The Spitzer Large Area Survey with Hyper-Suprime-Cam (SPLASH; e.g., \citealt{steinhardt2014}) which offers us robust $M_{\rm star}$ measurements at the redshift range of our ALPINE sample. 
The details of the physical properties and the ancillary data sets for our ALPINE sources are presented in \cite{faisst2020} and \cite{schaerer2020}.

\begin{figure*}
\begin{center}
\includegraphics[trim=0cm 0cm 0cm 0cm, clip, angle=0,width=1.0\textwidth]{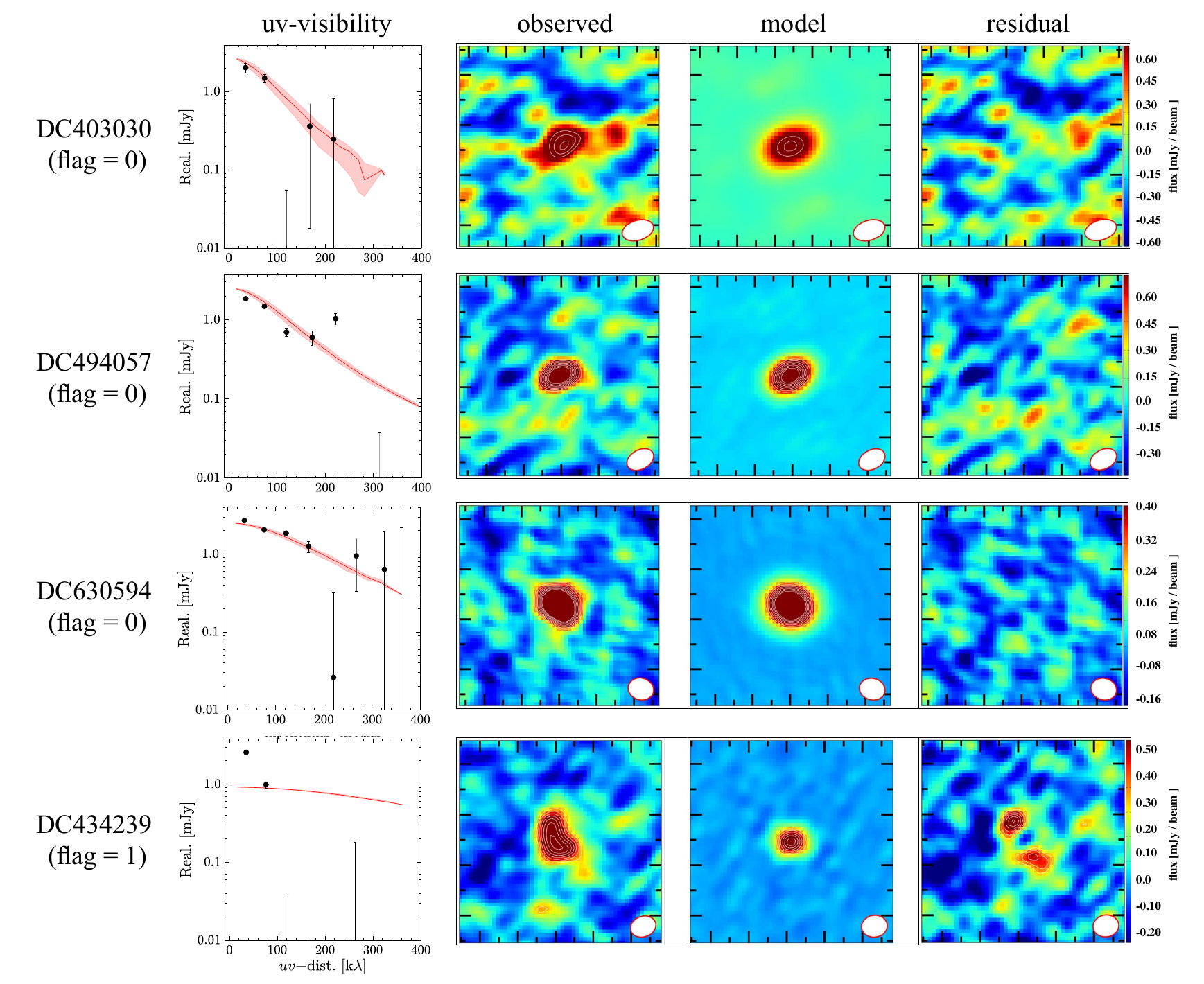}
 \caption[]{
Examples of the \cii\ size measurements for four ALPINE sources using {\sc uvmultifit}.
{\bf Left}: 
The best-fit model (red line and region of 1$\sigma$ uncertainty) to the $uv$-visibility data. 
Black circle denotes the median of the visibility-averaged data in bins of 50 ${\rm k}\lambda$, where the error bar shows the standard error of median. 
{\bf Right}:
Natural-weighted $8''\times8''$ velocity-integrated maps of the \cii\ line emission. 
The observed, best-fit model, and residual (= observed -- model) maps are presented in the left to right panels. 
White contours are drawn at 1$\sigma$ intervals starting from the 2$\sigma$ level.
The ALMA synthesized beam is shown in the bottom right corner. 
The flag values (0: reliable, 1: unreliable) are presented in the parenthesis. 
\label{fig:uv-amp}}
\end{center}
\end{figure*}

\subsection{Data Products}
\label{sec:product}

The details of data calibration, reduction, and \cii\ line extractions are described in \cite{bethermin2020}. 
Here we briefly overview the data products. 
The data were reduced with the Common Astronomy Software Applications package (CASA; \citealt{mcmullin2007}). 
The major steps were performed with the scripts provided by the ALMA Observatory. 
We applied additional flags, especially for a few bad antennas with suspicious behaviors \citep[see][]{bethermin2020}. 
With the CASA task {\sc tclean}, continuum maps were produced by utilizing the line-free channels in all spectral windows, 
while \cii\ line cubes were generated after continuum subtraction from the observed visibility data with {\sc uvcontsub}.  
The {\sc tclean} routines were executed down to the 3$\sigma$ level with a maximum iteration number of 500. 
We adopted a natural weighting for imaging with a pixel scale of $0.''15$ and a common spectral channel bin of 25 km s$^{-1}$, 
achieving a typical angular resolution $0.''9$ and a sensitivity of 0.35 mJy beam$^{-1}$ per spectral channel.  

We produced velocity-integrated (moment 0) \cii\ maps using an interactive approach 
by allowing slight spatial offsets ($<1''$) from the rest-frame UV centroids 
to maximize the \cii\ line detection. We corrected the HST astrometry in advance using the GAIA DR2 catalog \citep{faisst2020}. 
For the optimized velocity-integrated \cii\ line maps, 
we identified 75 \cii\ lines from our primary ALPINE targets with a signal-to-noise ratio (SNR) above 3.5. 
{Since the noise fluctuation may affect the size measurements in low SNR maps,}   
we use 46 out of the 75 \cii\ line sources whose peak counts in the velocity-integrated maps exceed the SNR of 5 
in the following analyses to obtain reliable size measurements. 
These 46 sources are listed in Table \ref{tab:catalog}. 

Visual inspection was performed by seven independent classifiers using 
the continuum-subtracted ALMA 3D \cii\ data cubes and ancillary data. 
The \cii\ sources are classified into the following five galaxy types: 
1) rotator, 2) pair-merger (major or minor), 3) dispersion-dominated, extended, 4) dispersion-dominated, compact, and 5) weak.
Combining the classifications from the seven participants, we determine a final galaxy type based on the mode of the distribution. 
The details of the classification procedure are described in Le F\`evre et al. (2019). We provide the final galaxy type for our ALPINE sources in Table \ref{tab:catalog}. 


\begin{table*}
{\scriptsize
\caption{Our ALPINE Source Catalog (SNR $\geq5$)}
\label{tab:catalog}
\begin{tabular}{lcccccccccc}
\hline
\hline
Name  & $z_{\rm [CII]}$ & SNR & $r_{\rm e, [CII]}$ & flag$_{\rm [CII]}$ & $r_{\rm e, f814w}$ & flag$_{\rm f814w}$ & $r_{\rm e, f160w}$ & flag$_{\rm f160w}$ & morph. class & Halo class \\
      &                  &     &          (kpc)      &                     &       (kpc)         &                     &       (kpc)         &                     &              &            \\
(1)   &    (2)           & (3) &            (4)      &     (5)             &    	(6)          &           (7)       &         (8)         &  (9)                 &  (10)        &    (11)    \\
\hline
CG32 & 4.4105 & 12.4 & 1.94 $\pm$ 0.3 & 0 & 0.91 $\pm$ 0.13 & 1 & 1.47 $\pm$ 0.04 & 1 & 2 &  --  \\
DC308643 & 4.5253 & 7.4 & 1.25 $\pm$ 0.55 & 0 & 0.90 $\pm$ 0.15 & 1 & No data & -- & 2 &  --  \\
DC351640 & 5.7058 & 5.4 & 2.05 $\pm$ 0.95 & 0 & 0.24 $\pm$ 0.30 & 1 & No data & -- & 3 & b \\
DC372292 & 5.1364 & 9.9 & 0.74 $\pm$ 0.46 & 0 & 1.01 $\pm$ 0.46 & 1 & No data & -- & 2 &  --  \\
DC396844 & 4.5424 & 12.3 & 2.56 $\pm$ 0.33 & 0 & 0.58 $\pm$ 0.20 & 0 & No data & -- & 1 & a \\
DC403030 & 4.5594 & 5.1 & 3.23 $\pm$ 0.95 & 0 & 0.0 $\pm$ 655.91 & 1 & No data & -- & 2 &  --  \\
DC416105 & 5.6309 & 5.5 & 1.02 $\pm$ 0.86 & 1 & 0.97 $\pm$ 0.34 & 0 & No data & -- & 1 & b \\
DC417567 & 5.67 & 6.5 & 2.07 $\pm$ 0.58 & 0 & 0.65 $\pm$ 0.20 & 0 & 0.84 $\pm$ 0.13 & 0 & 2 &  --  \\
DC422677 & 4.4381 & 7.1 & 1.10 $\pm$ 0.5 & 1 & 0.58 $\pm$ 0.14 & 0 & No data & -- & 2 &  --  \\
DC432340 & 4.4045 & 5.8 & 0.72 $\pm$ 6.26 & 1 & 1.00 $\pm$ 0.30 & 0 & No data & -- & 2 &  --  \\
DC434239 & 4.4883 & 8.0 & 0.62 $\pm$ 24.89 & 1 & 1.94 $\pm$ 0.95 & 1 & No data & -- & 2 &  --  \\
DC454608 & 4.5834 & 6.4 & 2.95 $\pm$ 0.72 & 0 & 0.87 $\pm$ 0.30 & 0 & No data & -- & 2 &  --  \\
DC488399 & 5.6704 & 27.7 & 1.32 $\pm$ 0.16 & 0 & 0.47 $\pm$ 0.32 & 1 & No data & -- & 3 & c \\
DC493583 & 4.5134 & 8.5 & 1.89 $\pm$ 0.51 & 0 & 0.64 $\pm$ 0.17 & 1 & 1.09 $\pm$ 0.32 & 0 & 2 &  --  \\
DC494057 & 5.5448 & 17.5 & 2.48 $\pm$ 0.25 & 0 & 0.59 $\pm$ 0.17 & 0 & 0.88 $\pm$ 0.16 & 1 & 1 &  --  \\
DC494763 & 5.2337 & 11.0 & 1.12 $\pm$ 0.36 & 0 & 0.65 $\pm$ 0.24 & 0 & 0.81 $\pm$ 0.28 & 0 & 3 &  --  \\
DC519281 & 5.5759 & 7.0 & 0.36 $\pm$ 1.28 & 1 & 0.49 $\pm$ 0.19 & 0 & 0.79 $\pm$ 0.18 & 0 & 2 &  --  \\
DC536534 & 5.6886 & 5.1 & 4.16 $\pm$ 1.04 & 1 & 0.87 $\pm$ 0.35 & 0 & 1.08 $\pm$ 0.24 & 1 & 2 &  --  \\
DC539609 & 5.1818 & 9.1 & 1.65 $\pm$ 0.43 & 0 & 0.77 $\pm$ 0.16 & 0 & 1.11 $\pm$ 0.32 & 0 & 1 & b \\
DC552206 & 5.5016 & 15.4 & 3.41 $\pm$ 0.35 & 1 & 0.96 $\pm$ 0.64 & 1 & No data & -- & 2 &  --  \\
DC627939 & 4.5341 & 13.3 & 2.12 $\pm$ 0.31 & 0 & 0.86 $\pm$ 0.26 & 1 & No data & -- & 2 &  --  \\
DC630594 & 4.4403 & 11.1 & 1.70 $\pm$ 0.34 & 0 & 0.77 $\pm$ 0.19 & 0 & 1.04 $\pm$ 0.28 & 1 & 3 & a \\
DC683613 & 5.542 & 14.1 & 1.82 $\pm$ 0.33 & 0 & 0.57 $\pm$ 0.24 & 0 & 0.81 $\pm$ 0.21 & 1 & 3 & a \\
DC709575 & 4.4121 & 5.6 & 1.34 $\pm$ 1.62 & 1 & 0.72 $\pm$ 0.17 & 0 & 0.96 $\pm$ 0.22 & 0 & 1 & b \\
DC733857 & 4.5445 & 7.5 & 1.80 $\pm$ 0.58 & 1 & 0.60 $\pm$ 0.13 & 0 & No data & -- & 3 & b \\
DC773957 & 5.6773 & 8.6 & 2.68 $\pm$ 0.54 & 0 & 0.76 $\pm$ 0.53 & 1 & No data & -- & 2 &  --  \\
DC818760$^{\dagger}$ & 4.5613 & 26.1 & 2.59 $\pm$ 0.16 & 1 & 0.75 $\pm$ 0.17 & 1 & 1.07 $\pm$ 0.21 & 0 & 2 &  --  \\
DC834764 & 4.5058 & 5.7 & 3.17 $\pm$ 0.89 & 0 & 0.88 $\pm$ 0.15 & 1 & No data & -- & 3 &  --  \\
DC842313 & 4.5537 & 6.0 & 0.79 $\pm$ 8.82 & 1 & 1.90 $\pm$ 0.31 & 1 & 6.81 $\pm$ 0.92 & 1 & 2 &  --  \\
DC848185 & 5.2931 & 18.1 & 3.47 $\pm$ 0.25 & 1 & 0.90 $\pm$ 0.30 & 1 & 1.24 $\pm$ 0.18 & 1 & 3 &  --  \\
DC873321 & 5.1542 & 7.8 & 0.67 $\pm$ 17.97 & 1 & 0.91 $\pm$ 0.27 & 1 & 1.30 $\pm$ 0.26 & 0 & 2 &  --  \\
DC873756 & 4.5457 & 34.1 & 2.36 $\pm$ 0.11 & 0 & 1.08 $\pm$ 0.43 & 0 & 1.14 $\pm$ 0.39 & 0 & 2 &  --  \\
DC880016 & 4.5415 & 8.7 & 2.41 $\pm$ 0.50 & 0 & 0.77 $\pm$ 0.28 & 0 & No data & -- & 3 & a \\
DC881725 & 4.5777 & 12.4 & 2.26 $\pm$ 0.33 & 0 & 0.67 $\pm$ 0.21 & 0 & 0.94 $\pm$ 0.30 & 0 & 1 & a \\
VC5100537582 & 4.5501 & 8.1 & 1.73 $\pm$ 0.53 & 0 & 0.54 $\pm$ 0.18 & 1 & No data & -- & 3 & a \\
VC5100541407 & 4.563 & 12.0 & 4.85 $\pm$ 0.59 & 1 & 0.84 $\pm$ 0.41 & 0 & 1.39 $\pm$ 0.25 & 1 & 2 &  --  \\
VC5100559223 & 4.5627 & 6.4 & 2.86 $\pm$ 0.77 & 0 & 0.97 $\pm$ 0.43 & 0 & 0.86 $\pm$ 0.43 & 1 & 3 &  --  \\
VC5100822662 & 4.5205 & 12.8 & 2.59 $\pm$ 0.37 & 0 & 1.32 $\pm$ 0.33 & 0 & 0.97 $\pm$ 0.29 & 1 & 2 &  --  \\
VC5100969402 & 4.5785 & 11.0 & 1.62 $\pm$ 0.33 & 0 & 0.59 $\pm$ 0.15 & 0 & No data & -- & 3 &  --  \\
VC5100994794 & 4.5802 & 12.2 & 1.86 $\pm$ 0.32 & 0 & 1.63 $\pm$ 0.35 & 1 & 1.23 $\pm$ 0.38 & 1 & 3 &  --  \\
VC5101218326 & 4.5739 & 28.1 & 2.37 $\pm$ 0.15 & 0 & 1.46 $\pm$ 0.32 & 1 & No data & -- & 3 &  --  \\
VC510596653 & 4.5681 & 5.9 & 1.79 $\pm$ 0.76 & 0 & 0.68 $\pm$ 0.29 & 0 & 0.39 $\pm$ 2.43 & 1 & 3 & b \\
VC510786441 & 4.4635 & 9.3 & 2.94 $\pm$ 0.52 & 0 & 0.85 $\pm$ 0.13 & 0 & 0.91 $\pm$ 0.26 & 0 & 2 &  --  \\
VC5110377875 & 4.5505 & 18.9 & 2.63 $\pm$ 0.23 & 0 & 0.93 $\pm$ 0.30 & 0 & No data & -- & 1 & a \\
VC5180966608 & 4.5296 & 13.5 & 5.10 $\pm$ 0.42 & 1 & 0.71 $\pm$ 8.87 & 1 & 0.59 $\pm$ 0.24 & 0 & 2 &  --  \\
VE530029038 & 4.4298 & 6.9 & 2.76 $\pm$ 0.65 & 0 & 1.96 $\pm$ 0.14 & 1 & 2.90 $\pm$ 0.08 & 1 & 1 &  --  \\
\hline
\end{tabular}
\tablecomments{
\footnotesize{
(1) ALPINE source name. We refer to CANDELS$\_$GOODS, DEIMOS$\_$COSMOS, VUDS$\_$COSMOS, and VUDS$\_$ECDFS in \cite{lefevre2019} as CG, DC, VC, and VE, respectively.   
We list only 46 ALPINE sources with SNR of the \cii\ line $\geq$ 5 that are used for the \cii\ line size measurement in this paper. 
The entire sample of 118 ALPINE sources with the coordinate is presented in \citealt{lefevre2019}. 
(2) Spectroscopic redshift estimated from the \cii\ 158$\mu$m line.  
(3) The peak SNR in the velocity-integrated map. 
(4) Circularized effective radius of the \cii\ line emission measured with {\sc uvmultifit} (see text).  
(5) Flag for the reliability of the {\sc uvmultifit} fitting.  
We define flag $=0$ and 1 as a source whose fitting result is reliable and bad, respectively. 
(6) Circularized effective radius of the rest-frame UV emission in the HST/F814W map measured with {\sc galfit} (see text). 
(7) Flag for the reliability of the {\sc galfit} fitting for the F814W map. The same flag definition as (5). 
(8) Circularized effective radius of the rest-frame UV emission in the HST/F160W map measured with {\sc galfit} (see text). 
(9) Flag for the reliability of the {\sc galfit} fitting for the F160W map.  
The same flag definition as (5). 
(10) Galaxy type based on the morphology+kinematic classification (1: rotator, 2: pair-merger, 3: dispersion-dominated, extended, 4: dispersion-dominated, compact, 5: weak; see \citealt{lefevre2019}). 
(11) Halo type classification (a: \cii\ Halo, b: w/o \cii\ Halo, c: \cii\ \&\ Dust Halo; see text).  }}
}
$\dagger$ The detail properties are presented in \cite{jones2020}. \\
\end{table*}

\section{Data analysis}
\label{sec:analysis}

\subsection{ALMA Size Measurements}
\label{sec:alma_size}

Prior to measuring the \cii\ sizes of the ALPINE sources, we extract from the continuum-subtracted visibility data the channels containing \cii\ emission, referred to as the \cii\ line visibility. 
We then produce \cii\ line maps with the CASA {\sc tclean} task. Initial values of the \cii\ flux density and sizes are determined through the image-based fitting with the {\sc imfit} task over an area of $4''\times4''$. 
Next, we measure sizes in the \cii\ line visibility using {\sc uvmultifit} (\citealt{marti2014}), a tool for simultaneous fitting multiple objects in the visibility plane. 
We model the light profiles using a S$\acute{\rm e}$rsic function, assuming a fixed S$\acute{\rm e}$rsic index n=1 (i.e., exponential-disk profile),
since the spatial distribution of the \cii\ line is likely to be related to the gas distribution 
\citep[e.g.,][]{wolfire2003,vallini2015} generally in an exponential disk-like structure \citep[e.g.,][]{bigiel2012}. 
We fix the source center as estimated from {\sc imfit} to improve the convergence of the free parameters. 
We obtain the FWHM along the major axis, FWHM$_{\rm major}$ and the axis ratio. 
The value of the FWHM$_{\rm major}$ is converted to the effective radius $r_{\rm e, major}$ via the relation of 
FWHM$_{\rm major}= 0.826\times r_{\rm e, major}$
in the case of $n=1$ \citep{macarthur2003}, 
and $r_{\rm e, major}$ is also converted to the ``circularized'' radius $r_{\rm e}$ through $r_{\rm e}\equiv a\sqrt{b/a} = r_{\rm e, major}\sqrt{q}$, where $a$, $b$, and $q$ are the major, minor axes, and axis ratio, respectively. 
Hereafter, the $r_{\rm e}$ value is regarded as our size measurement. 
We confirm that a different model in S$\acute{\rm e}$rsic index of $n=0.5$ returns consistent  $r_{\rm e}$ values within $\sim 5\%$.

Figure \ref{fig:uv-amp} presents examples of the \cii\ line visibility with our best-fit results. 
The phase centers of the visibility are shifted to the center of the \cii\ source. 
As a complementary check for the visibility fitting quality in the image plane, 
Figure \ref{fig:uv-amp} also shows the observed, best-fit model, and residual ($=$ observed -- model) maps for the \cii\ line emission. 
If the best-fit visibility has a large offset from the data and/or the residual map shows clear negative and/or positive residuals (e.g., DC434239), 
we flag these objects as unreliable fitting results that are removed from the following analyses. 
We obtain the reliable (flag = 0) and unreliable (flag = 1) fitting results for 31 and 15 ALPINE sources, respectively. 
The majority of these 15 sources are classified as mergers (Section \ref{sec:product}), indicating that the unreliable fitting results are mostly caused by the complicated morphology in merging systems. 
In Table \ref{tab:catalog}, we summarize our size measurements for the \cii\ line $r_{\rm e, [CII]}$ together with their associated flags. 

\begin{figure*}
\begin{center}
\includegraphics[trim=0cm 0cm 0cm 0cm, clip, angle=0,width=1.0\textwidth]{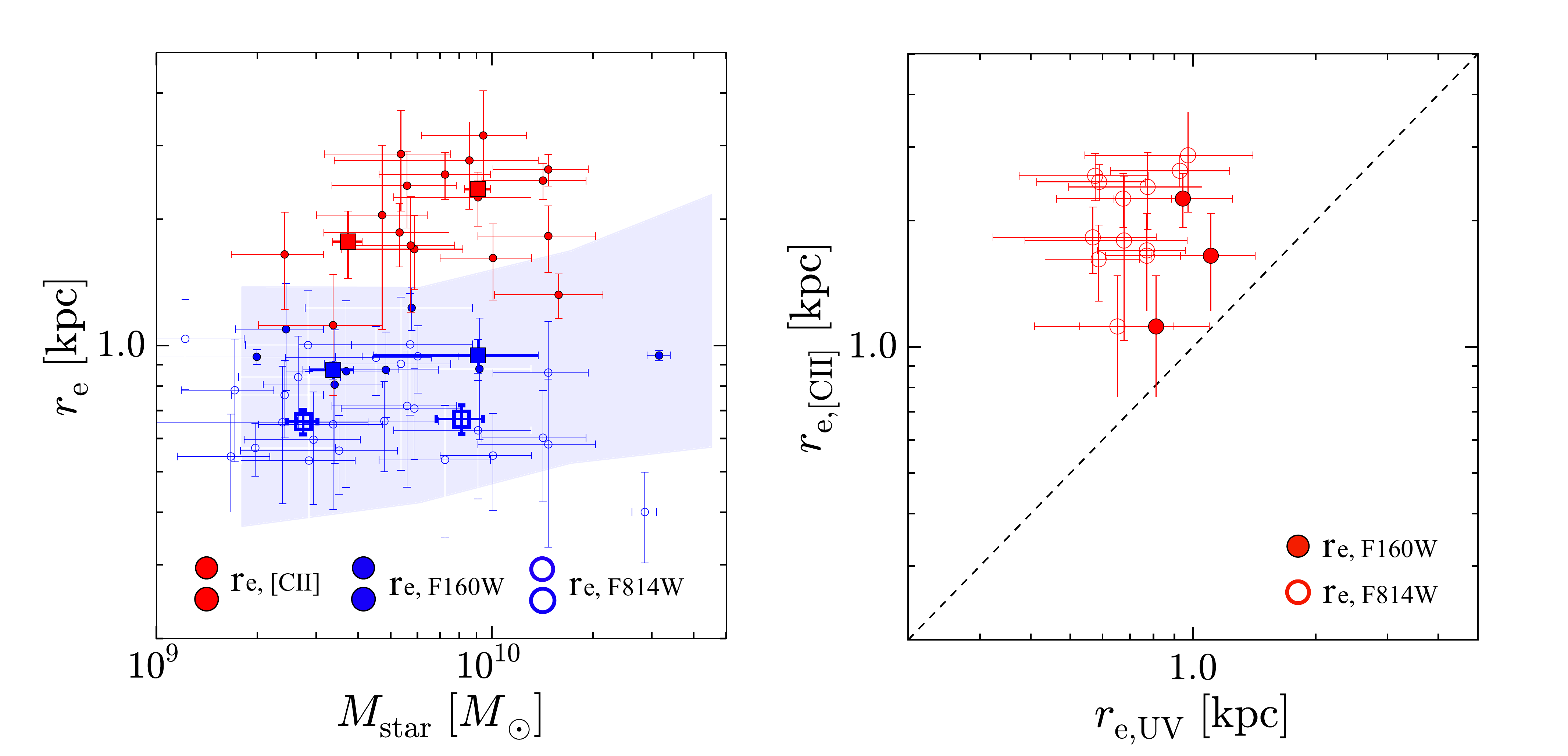}
 \caption[]{
Physical sizes of ALPINE sources. 
{\bf Left:} 
Size versus stellar mass. 
Red and blue circles indicate our \cii\ and rest-frame UV size measurements as a function of $M_{\rm star}$. 
The square symbols represent the median values in bins of $M_{\rm star}$, where the error bars denote the standard error of the median. 
We present the reliable size measurement results (flag = 0) of the \cii\ line and rest-frame UV for the 19 and 32 and ALPINE sources, respectively, 
that are not classified as mergers.
Both F814W and F160W measurements are shown for 6 ALPINE sources. 
Blue open and filled symbols denote the rest-frame UV size measurements in the HST/F814W and F160W maps, respectively. 
The blue shaded region gives the 1$\sigma$ range of the rest-frame UV size distribution of $\sim$4,000 normal star-forming galaxies presented in \citet{shibuya2015}. 
{\bf Right:} Comparison between \cii\ and rest-frame UV sizes among the individual galaxies. 
The red open and filled circles indicate whether the rest-frame UV sizes are estimated from the HST/F814W and F160W maps, respectively.
We present the \cii\ and rest-frame UV size measurements for 12 ALPINE sources whose both measurements are reliably (flag = 0) obtained. 
Both F814W and F160W measurements are shown for 3 ALPINE sources. 
\label{fig:cii-uv_size}}
\end{center}
\end{figure*}

\subsection{HST Size Measurements}
\label{sec:hst_size}

To compare the \cii\ sizes with the stellar distribution, we measure the rest-frame UV size of our ALPINE sources based on the high spatial resolution images of HST ACS (F814W image in most cases; \citealt{scoville2007,koekemoer2007}).
If the source has been observed with WFC3/IR, we also use the F160W image \citep{koekemoer2011,faisst2020}. 
Since the redshifts of our ALPINE sources fall in the range of $z\sim 4.5-5.5$, the rest-frame wavelengths correspond to $\sim$1300--1500 ${\rm \AA}$ and 2500--2900 ${\rm \AA}$ for the F814W and F160W images with the point-spread function (PSF) FWHM sizes of $0.''09$ and $0.''18$, respectively. 
For size measurements, we extract $5''\times5''$ cutout images from the F814W and F160W maps around ALPINE targets. 
The cutout size is sufficiently large to evaluate the entire galaxy structure. 
Based on available photometric redshift $z_{\rm ph}$ catalogs \citep{laigle2016, momcheva2016}, 
we model foreground objects ($z_{\rm ph} \lesssim2$) with GALFIT \citep{peng2010} and remove them from the HST images, 
if the objects are identified within a radius of $1.''0$ from the ALPINE sources. 
In addition to the 46 ALPINE sources used in the \cii\ line size measurements, here we also use 29 ALPINE sources whose \cii\ line emission are detected with 3.5 $\leq$ SNR $<$ 5 to obtain statistical results.  
We list these 29 ALPINE sources in Table \ref{tab:catalog2}. 

We measure the rest-frame UV sizes in the same manner as previous high-$z$ galaxy studies  in the parametric approach \citep[e.g.,][]{vanderwel2012, ono2013, shibuya2015} 
based on two-dimensional (2D) surface-brightness (SB) profile fitting with {\sc galfit} \citep{peng2010}. 
Note that we do not adopt the non-parametric approach \citep[e.g.,][]{ribeiro2016} for the rest-frame UV size measurement. 
This is because the non-parametric approach largely depends on the data properties (e.g., depth and resolution), 
and our goal is to compare the rest-frame UV size to that of the \cii\ line which is obtained from the ALMA data sets having very different data properties from the HST data. 
We fit a single S$\acute{\rm e}$rsic profile \citep{sersic1963} to the 2D SB distribution of each galaxy.  
Here, we fix $n=1$ to perform self-consistent measurements and a fair comparison with the \cii\ size measurements 
for which we adopt an exponential-disk ($n=1$) profile (Section \ref{sec:alma_size}). 
We use PSF models that are provided by the 3D-HST project \citep{skelton2014}.
We obtain $r_{\rm e, major}$ and an axis ratio that is then converted to the circularized size $r_{\rm e}$ in the same manner as the \cii\ size measurements. 
If we find clear negative and/or positive peaks remaining in the residual maps or the divergence in the fitting process,
we flag the objects with these unreliable fitting results. 
From the 75 ($= 46 + 29$) ALPINE sources, 
we obtain the reliable fitting results for 52 ALPINE sources in either or both of the F814W and F160W images. 
We list our size measurements for the rest-frame UV emission $r_{\rm e, UV}$ together with their flags (reliable: flag $=0$, unreliable: flag $=1$) in Table \ref{tab:catalog} and Table \ref{tab:catalog2}.

\section{Results}
\label{sec:results} 

\subsection{[CII] and rest-frame UV sizes}
\label{sec:cii-uv_size}

In Figure \ref{fig:cii-uv_size}, we show the \cii\ and rest-frame UV sizes of the ALPINE sources. 
To minimize potential systematic errors from merging systems, we use the size measurement results only for ALPINE sources not classified as mergers (Section \ref{sec:product}). 

The left panel presents $r_{\rm e}$ as a function of $M_{\rm star}$, 
where the filled red, filled blue, and open blue circles denote $r_{\rm e, [CII]}$, $r_{\rm e, F160W}$, and $r_{\rm e, F814W}$, respectively for each galaxy. 
The square symbols are the median sizes in the bins of $M_{\rm star}$. 
We find the median value to be  $r_{\rm e, [CII]} = 2.1 \pm 0.16$ kpc, $r_{\rm e, F160W} = 0.91\pm0.06$ kpc, and $r_{\rm e, F814W} = 0.66\pm 0.04$ kpc 
with a standard error on the median.  
The \cii\ sizes are thus larger than the rest-frame UV sizes by factors of $\sim2-3$ on a statistical basis.
We also find that our rest-frame UV sizes are in agreement with published studies such as those of \cite{shibuya2015} (blue shaded region in Figure \ref{fig:cii-uv_size}) based on an analysis of $\sim$4,000 normal star-forming galaxies at $z=4-6$ and an equivalent S$\acute{\rm e}$rsic profile fitting procedure. 
This confirms that our ALPINE sources, selected from main-sequence galaxies, represent the typical star-forming galaxy population at $z=4-6$. 
These results indicate that the condition of $r_{\rm e, [CII]} > r_{\rm e, UV}$ is a general physical characteristic among the high-$z$ star-forming galaxies at $z=4-6$ in the $M_{\rm star}$ range of $\sim2\times10^{9-10}\,M_{\odot}$. 

In the right panel of Figure \ref{fig:cii-uv_size}, 
we show a comparison between $r_{\rm e, [CII]}$ and $r_{\rm e, UV}$ for the 12 ALPINE sources whose \cii\ and rest-frame UV size measurements are reliable (flag = 0). 
We find that all sources fall in the area of $r_{\rm e, [CII]} > r_{\rm e, UV}$ mostly higher than the individual errors, 
which is consistent with previous results of \cite{carniani2018b} for star-forming galaxies at $z=5-7$ including the mergers. 
From both statistical and individual results, 
we conclude that the \cii-line emitting regions are generally more extended than the rest-frame UV emitting regions in the high-$z$ star-forming galaxies at $z=4-6$. 
This is consistent with the recent simulation results \citep[e.g.,][]{vallini2015,pallottini2019}.
 
\begin{figure}
\begin{center}
\includegraphics[trim=0cm 0cm 0cm 0cm, clip, angle=0,width=0.5\textwidth]{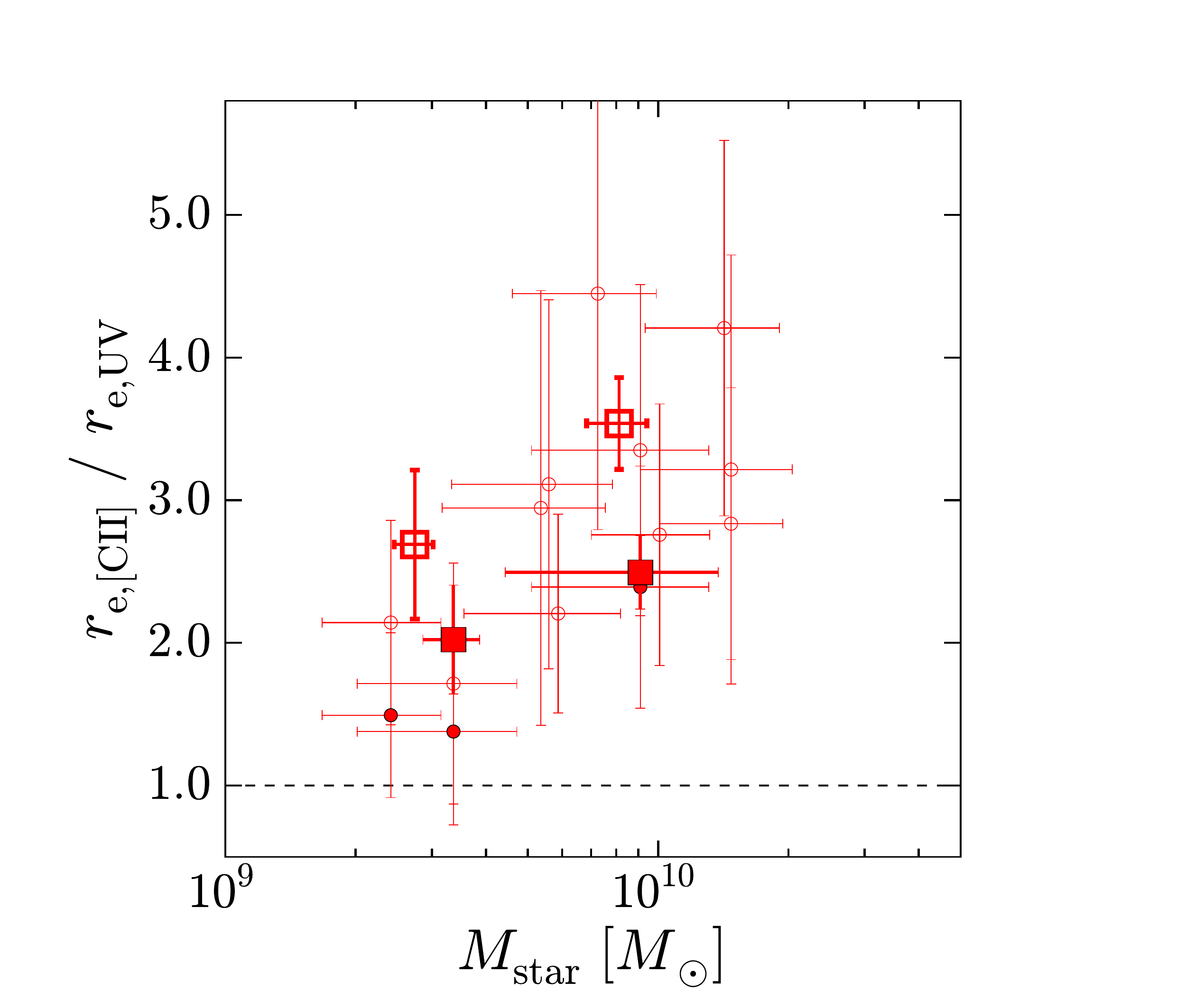}
\caption[]{Ratio of $r_{\rm e, \, [CII]}$ to $r_{\rm e, \,UV}$ as a function of $M_{\rm star}$. Red open and filled circles are defined as in the right panel of Figure \ref{fig:cii-uv_size}. Median values (squares) are the same as presented in the left panel of Figure \ref{fig:cii-uv_size}.
The rest-frame UV sizes shown with the open and filled squares are estimated from the HST/F814W and F160W maps, respectively.  
Spearman's rank test shows the significance of the positive correlation at $\sim$98\%. 
\label{fig:cii-uv_ratio}}
\end{center}
\end{figure}

In Figure \ref{fig:cii-uv_ratio}, 
we display the ratio of $r_{\rm e, [CII]}$ to $r_{\rm e, UV}$ as a function of $M_{\rm star}$. 
The square and circle plots are obtained from the median in the $M_{\rm star}$ bins and the individual results in the left and the right panels of Figure \ref{fig:cii-uv_size}, respectively. 
We find that the ratio has a positive trend towards high values of $M_{\rm star}$. 
With the individual results, Spearman's rank test shows the significance of this positive trend at $\sim$98\%.
This may indicate that the physical origin of the extended \cii\ line structure is related to $M_{\rm star}$ or SFR, 
since our ALPINE sources are main-sequence galaxies (i.e., they lie on the  $M_{\rm star}$--SFR plane; Section \ref{sec:overview}).  
We discuss the details of the possible physical mechanisms of the extended \cii\ line structure in Section \ref{sec:discuss}. 

We note that the small size in the rest-frame UV emission may be contributed by the smaller PSFs of HST than that of ALMA.  
To test this hypothesis, we perform a Monte Carlo simulation for the rest-frame UV emission in the F814W map. 
We model 1000 artificial sources having the 2D exponential-disk profile with $r_{\rm e}$ of 2.1 kpc which is the typical value for the \cii\ line emission.  
The axis ratio and the position angles for the artificial sources are taken randomly in the ranges of 0.1 -- 1.0 and 0 -- 90 deg. 
We then add the random noise on the 2D map and convolve the artificial sources with the F814W PSF. 
We add the random noise until the SNR value in the convolved map becomes comparable to the average value ($\sim 15$) of the rest-frame UV emission among our ALPINE sources. 
We finally measure the $r_{\rm e}$ values for the artificial sources with {\sc galfit} in the same manner as our size measurement for the F814W maps. 
We obtain the median and the standard deviation of the $r_{\rm e}$ measurements of 2.15 kpc and 0.14 kpc for the 1000 artificial sources.  
This simulation indicates that the $r_{\rm UV}$ measurements should show the typical value of $\sim2$ kpc, 
if the rest-frame UV emission is also extended as much as the \cii\ line emission. 
The difference of the PSFs between HST and ALMA is thus not the cause of the relatively small size in the rest-frame UV emission compared to the \cii\ line emission. 

We also note that the $r_{\rm e}$ measurements in the F814W image are generally smaller than those in the F160W image (Figure \ref{fig:cii-uv_size} and Table \ref{tab:catalog}). 
In the similar Monte Carlo simulations, we confirm that this difference is not caused by the different PSFs, neither. 
This is probably because of the dust absorption which affects the spatial morphology more significantly at the rest-frame short wavelength of F814W rather than F160W.

\begin{figure*}[h]
\begin{center}
\includegraphics[trim=0cm 0cm 0cm 0cm, clip, angle=0,width=1.0\textwidth]{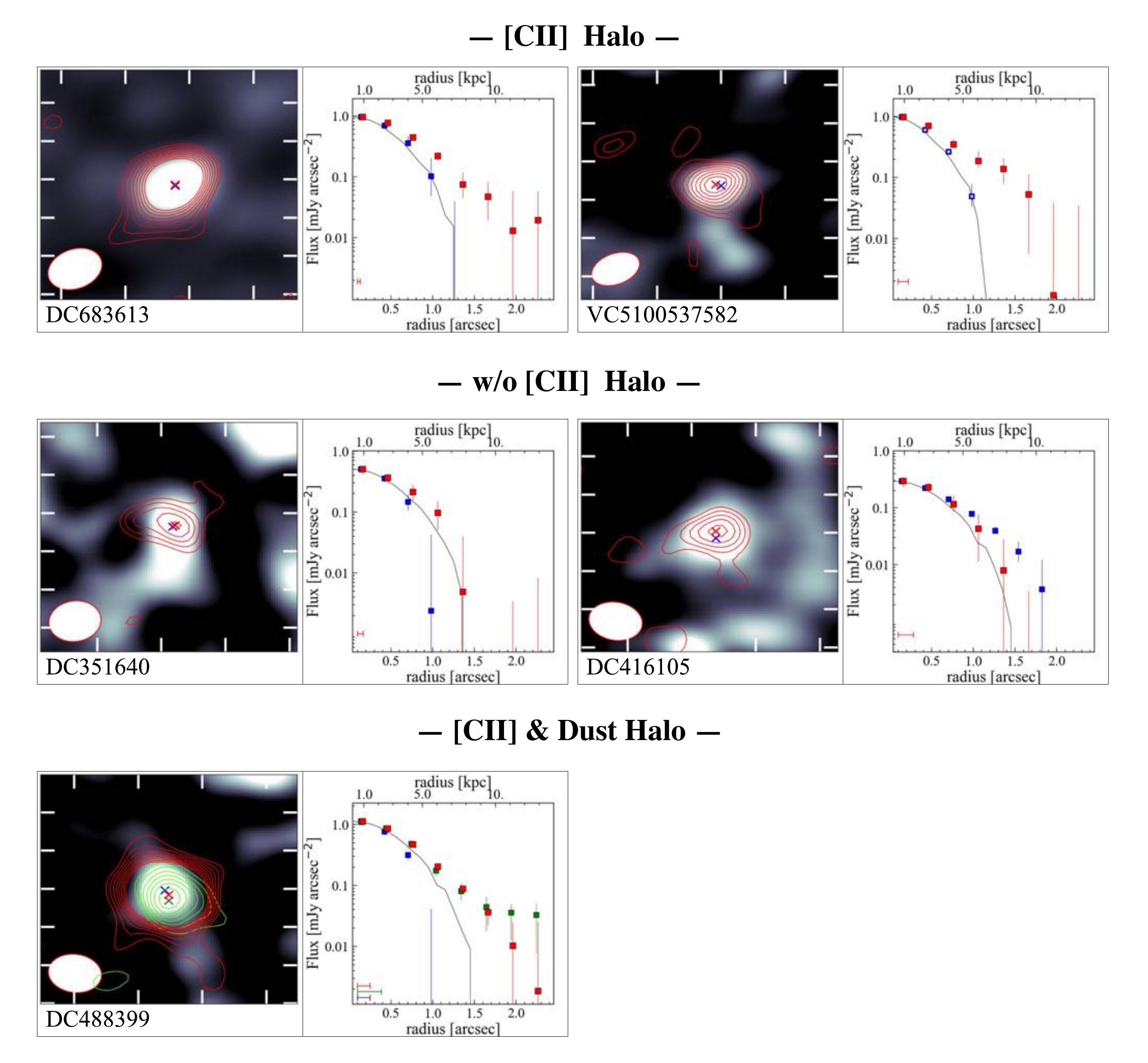}
 \caption[]{
Representative examples of the spatial distributions of the \cii\ line, the rest-frame FIR, and UV continuum. 
The remaining objects are presented in Figure \ref{fig:postage_ap}. 
By evaluating the extended components, we identify three categories: \cii\ Halo, w/o \cii\ Halo, and \cii\ \& Dust Halo objects (see text).　
{\bf Left}: 
The grey-scale ($6''\times6''$) image shows the rest-frame UV emission from the HST imaging \citep{koekemoer2007,koekemoer2011}. 
Red and green contours denote the 2.5$\sigma$, 3.5$\sigma$, 4.5$\sigma$, ..., 9.5$\sigma$ levels of the ALMA \cii\ line and the rest-frame FIR continuum emission, respectively. 
The green contour is presented only for the sources whose dust continuum emission is detected above the $5\sigma$ level.  
The HST image and ALMA contours are smoothed by the ALMA synthesized beam and the HST PSF, respectively, to match the spatial resolutions between the HST and ALMA maps. 
The smoothed PSF is shown at the bottom left. 
The red, green, and blue crosses mark the peak pixel positions of the \cii\ line, the rest-frame FIR and UV continuum, respectively.
{\bf Right}: 
Radial surface brightness profiles of the \cii\ line (red squares), the rest-frame FIR (green squares) and UV (blue squares) continuum. 
The black solid curve shows the ALMA synthesized beam. 
All radial profiles are estimated from the median value in an annulus with a width of $0.''3$,
and their peaks are normalized to the \cii\ one.
The errors denote the 16--84th percentile of the median values in 100 random annuli. 
The spatial offsets between UV -- [C {\sc ii}], UV -- FIR, and \cii\ -- FIR peaks are presented in the red, green, and black bars at the bottom left. 
For the HST maps with foreground objects removed, the radial profiles of the rest-frame UV continuum are shown by open blue squares. 
\label{fig:postage_radial}}
\end{center}
\end{figure*}

\subsection{\cii\ Halo around Individual Galaxies}
\label{sec:cii_halo}

\cite{fujimoto2019} report the existence of a 10-kpc scale \cii\ halo surrounding young galaxies via the visibility-based stacking of deep ALMA data for 18 star-forming galaxies at $z=5-7$. Using a similar method, the stacked images of ALPINE sources show a 15-kpc scale extended structure in the \cii\ line emission \citep{ginolfi2020}. Making full use of the large ALPINE and ancillary data sets, we investigate 
the spatial extents of the \cii\ line and the continuum down to the outskirts, instead of the $r_{\rm e}$ values, and test whether the existence of the \cii\ halo can be identified for individual galaxies.  
For this exercise, we focus on the 23 ALPINE sources whose \cii\ lines are detected above the 5$\sigma$ level and are not classified as mergers (Section \ref{sec:product}; see \citealt{lefevre2019}) 
For the sources with (without) the HST/F160W data, we use the F160W (F814W) data as the rest-frame UV image. 
To compare morphological properties between ALMA and HST maps, 
we match their resolutions. 
Since the PSF of ALMA (i.e., synthesized beam) is difficult to match to HST just by smoothing the HST PSF with a specific kernel, 
the ALMA (HST) maps are both convolved with the PSF of HST (ALMA) for matching their resolutions. 
Before convolution, we identify foreground interlopers ($z_{\rm ph}\lesssim 2$) around our ALPINE sources based on the photometric redshift catalogs \citep{laigle2016, momcheva2016} and remove these interlopers from the HST maps by modeling with {\sc galfit}. 

Figures \ref{fig:postage_radial} and \ref{fig:postage_ap} show the spatial distribution and the radial profiles of the \cii\ line, the rest-frame UV, 
and, when available, the rest-frame FIR continuum of our ALPINE sources. 
The HST astrometry is corrected with GAIA DR2 catalog \citep{faisst2020}. 
We find that the \cii\ line emitting region is generally more extended than the rest-frame UV and FIR continuum emitting regions.
In some cases, the \cii\ line extends up to radii of $\sim$ 10 -- 15 kpc, 
while both the rest-frame UV (and FIR) continuum drops off (e.g., DC683613, DC880016, VC5100537582). 
This indicates that the 10-kpc scale \cii\ halos are individually identified in a number of galaxies. 
However, in other cases, we find that the \cii\ lines do not show such extended morphology with the radial profile similar to the rest-frame UV and/or FIR continuum (e.g., DC351640, DC733857). 
This suggests that the \cii\ morphology is not always extended beyond the size of the stellar/dust emitting regions. 

\begin{table}
\caption{Subgroups of Isolated ALPINE Sources}
\label{tab:subgroup}
\begin{center}
\begin{tabular}{ccc}
\hline \hline
Halo class & criteria & Number \\
(1) &  (2) & (3) \\
\hline 
                                              &  $S_{\rm ext, \, [CII]} \geq 4\sigma$             &  \\ 
\cii\ Halo                           &  $S_{\rm ext,\, FIR} < 3\sigma $          &     7   \\
                                              & $S_{\rm ext, \,UV} < 3\sigma $           &     \\ \hline
w/o \cii\ Halo                   & $S_{\rm ext, \, [CII]} < 3\sigma$  & 6  \\         \hline
                                              &  $S_{\rm ext, \, [CII]} \geq 4\sigma$             &  \\ 
\cii\ \& Dust  Halo            &  $S_{\rm ext,\, FIR} \geq 4\sigma $          &    1    \\
                                              & $S_{\rm ext, \,UV} < 3\sigma $           &          \\
\hline
\end{tabular}
\end{center}
\tablecomments{
\footnotesize{
(1): Name of subgroups of the isolated ALPINE sources based on the properties of the extended emission beyond the ALMA beam. 
(2): Criteria of the subgroups. 
(3): Number of the sources that meet each criteria of the subgroups among the 23 ALPINE sources whose \cii\ lines are detected above the 5$\sigma$ level and are not classified as mergers. The rest of the marginal 9 ($= 23 - 7 - 6- 1$) ALPINE sources are not used for the discussion in Section \ref{sec:discuss}. 
}}
\end{table} 


To quantitatively identify the objects with a \cii\ halo whose structure is extended more than the continuum emission,
we evaluate the significance level for the spatially extended intensity of the \cii\ line emission ($S_{\rm ext, [CII] }$). 
Here we use the 23 ALPINE sources, not classified as mergers, whose \cii\ lines are detected above the 5$\sigma$ level. 
Using the velocity integrated \cii\ maps, we calculate the extended \cii\ line intensity in an aperture having a radius of 10 kpc and masking the emission in a central area corresponding to the ALMA synthesized beam. 
We estimate the error using a random aperture method. 
We detect 12 sources with an extended \cii\ line intensity above the $4\sigma$ level in the range of 4.1$\sigma -10.9\sigma$ levels. 
In the same manner, we also evaluate the significance level of the extended emission for the rest-frame FIR  ($S_{\rm ext, FIR} $) and UV continuum ($S_{\rm ext, UV }$). 
We identify 7 out of the 12 sources whose $S_{\rm ext, FIR} $  and  $S_{\rm ext, UV }$  are both below 3$\sigma$ level. 
We regard these 7 sources, DC396844, DC630594, DC683613, DC880016, DC881725, VC5100537582, and VC5110377875 as the ``[C {\sc ii}] Halo'' objects.  
This suggests that at least 30\% ($\sim$ 7/23) of the isolated ALPINE sources have an individual \cii\ halo at the depth of the current ALPINE data set. 

We also identify another 6 sources (DC351640, DC416105, DC539609, DC709575, DC733857, and VC510596653) 
with the extended \cii\ line intensity below the $3\sigma$ level that are unlikely to have the \cii\ halo structure. 
We refer to these 6 sources as  ``w/o [C {\sc ii}] Halo'' objects. 
We compare the physical properties of [C {\sc ii}] Halo and w/o [C {\sc ii}] Halo objects in Section \ref{sec:discuss}. 

We further highlight DC488399, a galaxy that uniquely has an extended morphology in the rest-frame FIR continuum and \cii\ line ($S_{\rm ext, FIR}$ and $S_{\rm ext, [CII]}$   $  > 4\sigma$), but a compact morphology in the rest-frame UV continuum ($S_{\rm ext, UV} < 3\sigma$).
Recent ALMA observations among high-$z$ star-forming galaxies reveal that the dusty, rest-frame FIR emitting region is generally more compact than the rest-frame UV and/or optical emitting regions \citep[e.g.,][]{simpson2015a, ikarashi2015, hodge2016, fujimoto2017, fujimoto2018}. 
For DC488399, the opposite trend is seen with the extended FIR component possibly co-spatial with the extended \cii\ line structure.
We refer to DC488399 as a ``[CII] \& Dust Halo'' object, and discuss its physical properties in Section \ref{sec:discuss}. 
We summarize the detection criteria and the number of the sources in these subgroups (\cii\ Halo, w/o \cii\ Halo, and Dust \& \cii\ Halo) in Table \ref{tab:subgroup}.

\subsection{Spatial Offsets ([CII], FIR, and UV)}
\label{sec:offset}

Recent observational and theoretical studies report a spatial offset between the centroids of the \cii\ line and the rest-frame UV continuum with a physical scale of $\sim4$ kpc \citep[e.g.,][]{vallini2013,maiolino2015,carniani2017}. 
These spatial offsets may affect the \cii\ line morphology compared to the rest-frame UV and FIR ones. 
Making full use of the large ALPINE and ancillary data sets again, 
we evaluate the spatial offsets among the \cii\ line, the rest-frame UV, and the rest-frame FIR continuum to check for any potential effects on the detection of an extended structure of the \cii\ line emission. 
To remove the obvious effects from merging galaxies, we focus on the 23 ALPINE sources analyzed in Section \ref{sec:cii_halo} that do not include the sources classified as mergers. 

In Figure \ref{fig:postage_radial} and \ref{fig:postage_ap}, we show the peak positions (left panels) as well as the offset scales (right panels) among these three emissions. To carry out a fair comparison, we measure the emission peaks at the peak pixel positions in the smoothed HST and ALMA maps whose resolutions are matched. 
For the 23 ALPINE sources, 
the median (standard deviation) values for the spatial offsets are estimated to be $0\farcs15$  ($0\farcs09$), $0\farcs25$  ($0\farcs08$), and $0\farcs15$  ($0\farcs12$) between \cii --UV, FIR--UV, and \cii --FIR, respectively. 
Although the peak pixel positions may depend on the pixel scale of the maps, we confirm that we obtain the consistent median and standard deviation of the spatial offsets in the smaller pixel scale by reproducing and reanalyzing the ALMA map with a pixel scale of $0\farcs03$.

The approximate positional accuracy of the ALMA map $\Delta p$ in milliarcsec is given by 
\begin{eqnarray}
\Delta p = \frac{70000}{\nu * B * \sigma},
\end{eqnarray}where $\sigma$ is the peak SNR in the map, $\nu$ is the observing frequency in GHz, and $B$ is the maximum baseline length in kilometers (ALMA technical handbook \footnote{Section 10.5.2: https://almascience.nao.ac.jp/documents-and-tools/cycle7/alma-technical-handbook}). 
Assuming the general property of the ALPINE data set ($B=0.2$  in the configuration of C43-1, $\nu\sim$ 330 GHz),  the ALPINE \cii\ line map with a peak SNR = 5 -- 10 has the $\Delta p $ value of $\sim0\farcs1-0\farcs2$. 
Moreover, even with the HST astrometry corrected using the GAIA DR2 source catalog, the uncertainty in the astrometry correction still remains at the $\sim0\farcs1$ scale due to the scatter in the source positions between the public HST and the GAIA DR2 catalogs (see Fig.15 of \citealt{faisst2020}). Thus, our measurements of spatial offsets are comparable to the uncertainties of the positional accuracy for the ALPINE sources and the HST astrometry.  

\begin{figure}
\begin{center}
\includegraphics[trim=0cm 0cm 0cm 0cm, clip, angle=0,width=0.5\textwidth]{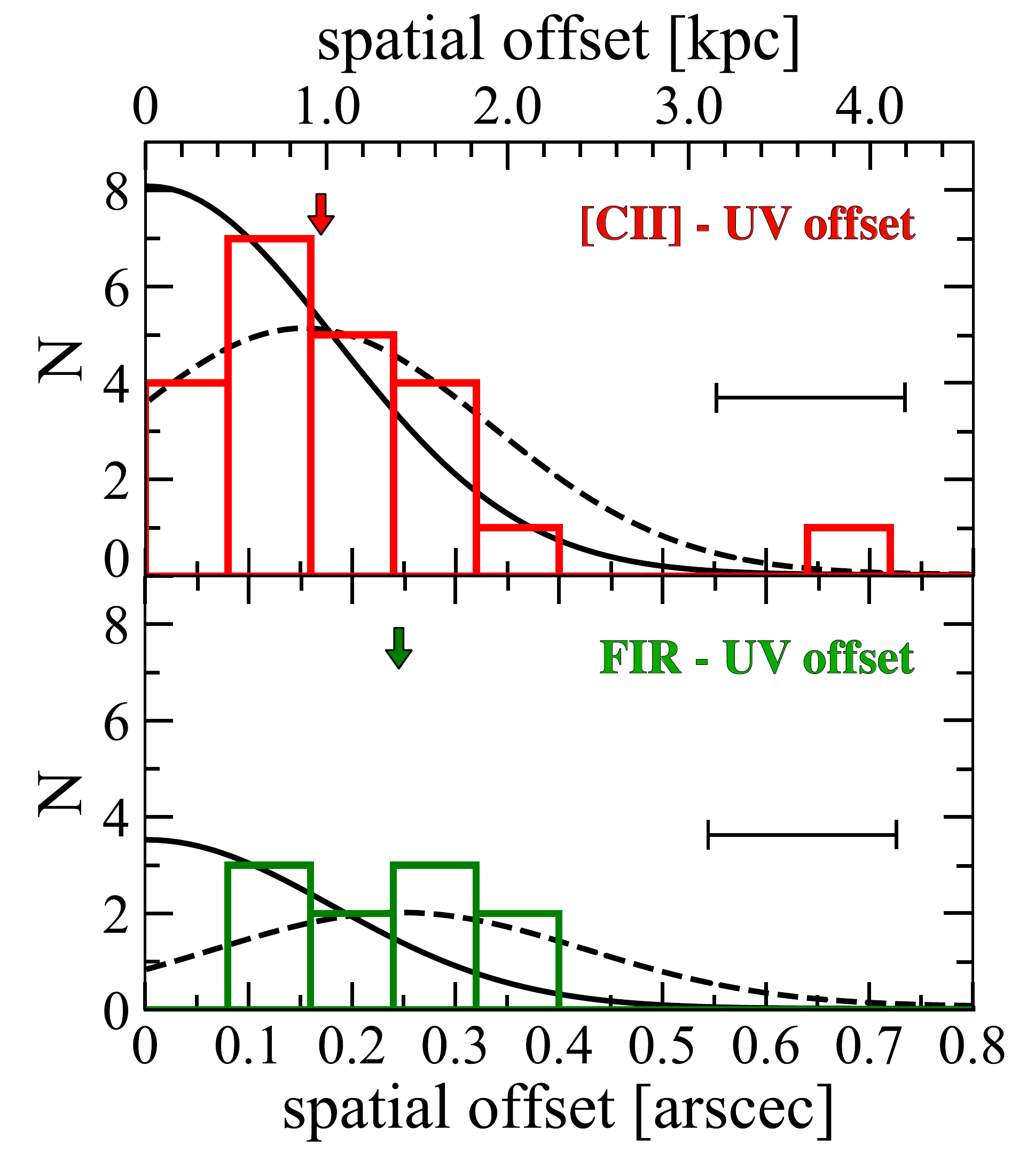}
 \caption[]{
Radial distributions of the spatial offset between \cii -- UV (top; red histogram) and FIR--UV (bottom; green histogram). 
The solid and dashed curves denote the Gaussian distributions in the cases of i) and ii), respectively (see text). 
The arrows indicate the median values of the spatial offsets. 
The horizontal bar shows the propagated error from the positional accuracy and the astrometry correction that we adopt for the standard deviation of the Gaussian distributions.  
\label{fig:offset_test}}
\end{center}
\end{figure}

To test whether the observed spatial offsets are ascribed to the above uncertainties, we further examine the radial distribution of the spatial offsets. 
In Figure \ref{fig:offset_test}, we present the radial distribution of the spatial offsets between \cii --UV (red histogram) and FIR--UV (green histogram). 
There are two possible cases to produce these radial distributions:
i) The spatial offsets are intrinsically $\sim$zero. Still, the uncertainties cause the non-zero distribution, which follows the Gaussian with the center of zero. 
ii) The emission peaks have intrinsic spatial offsets comparable to the median values, where the distribution follows another Gaussian with the center of the median value.
For comparison in Figure \ref{fig:offset_test}, we also show the radial distributions of the Gaussian in the cases of i) and ii) with the solid and dashed curves, respectively.  
We perform the Kolmogorov-Smirnov (KS) test between the radial distributions of the spatial offsets and the Gaussian. 
In the spatial offset between \cii --UV, we find that we cannot rule out the possibility that the radial distribution of the spatial offsets is produced from the Gaussian in both cases. 
In the spatial offset between FIR--UV, we obtain the same results as \cii --UV, although the sample number is insufficient for the KS test. 
We thus cannot draw definite conclusions from the current data, whether the observed spatial offsets are just caused by the uncertainties or the intrinsic spatial offsets with the uncertainties.  It is also possible that both cases are taking place. 
These results indicate that the [C {\sc ii}], the rest-frame FIR,  and the rest-frame UV emitting regions may not always be displaced from each other.  

Importantly, even if spatial offsets exist on the typical scale of the median values of $\sim0\farcs1$--$0\farcs2$, corresponding to $\simeq$ 1 kpc at $z=4-6$, 
the 10-kpc scale \cii\ halo is unlikely to be explained by these marginal offsets.  Previous stacking studies also show that adopting different stacking centers (e.g., peak positions of \cii\ line and the rest-frame UV emission)
do not change the extended structure in the \cii\ line rather than the continuum \citep{fujimoto2019,ginolfi2020}. 
This also suggests that the spatial offsets are sufficiently small compared to the \cii\ halo structure. 
We conclude that the spatial offsets have a negligible effect on the measurement of the \cii\ halo. 
The potential offset and its physical origin will be explored in a future ALPINE work. 

\section{Discussion}
\label{sec:discuss}

\begin{figure}
\begin{center}
\includegraphics[trim=0cm 0cm 0cm 0cm, clip, angle=0,width=0.5\textwidth]{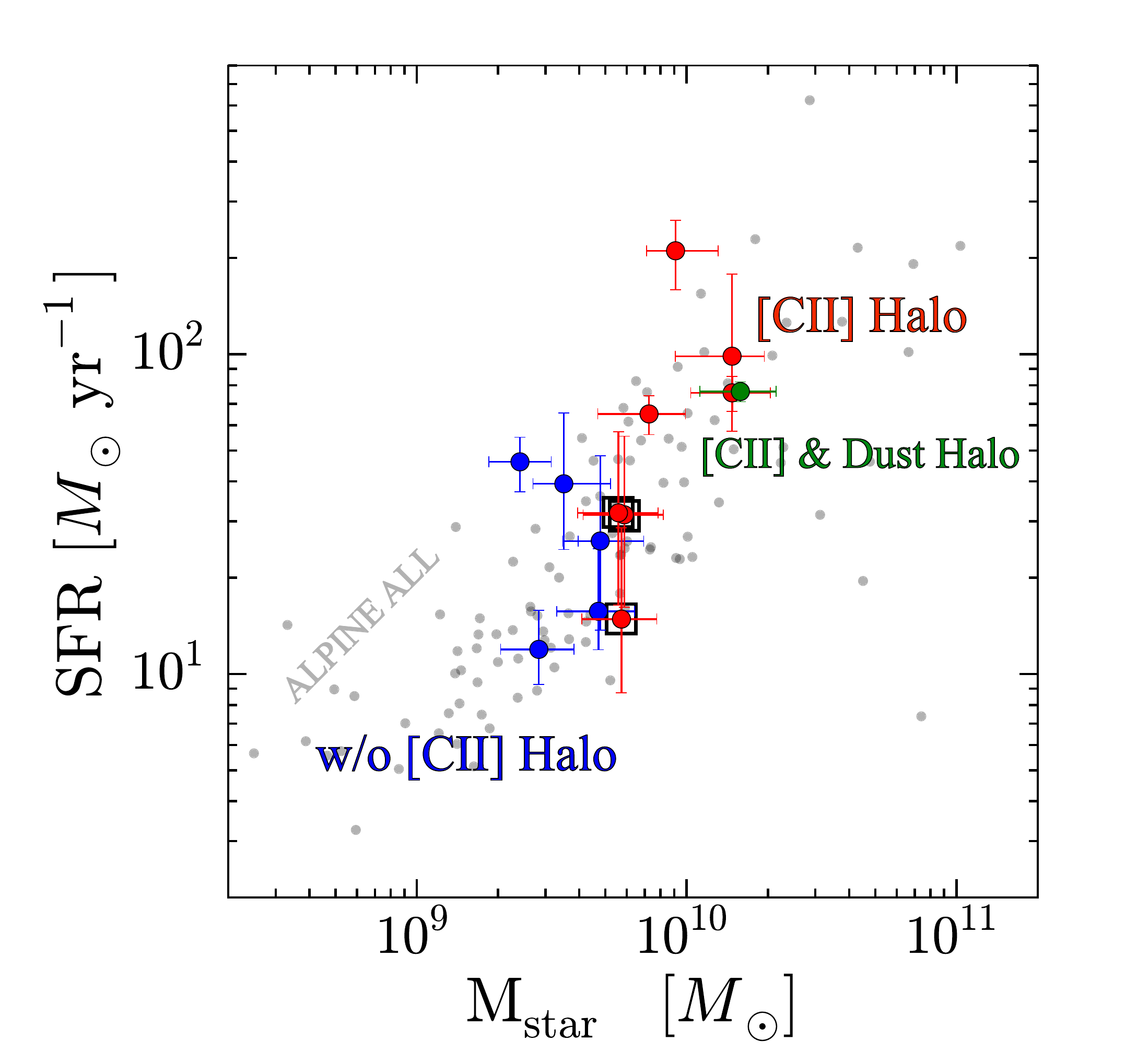}
 \caption[]{
SFR and $M_{\rm star}$ relation. 
The red and blue circles show the \cii\ Halo and w/o \cii\ Halo, respectively. 
Gray dots indicate the entire sample of the 118 ALPINE sources \citep{schaerer2020}. 
We mark three \cii\ Halo objects (DC630594, DC880016, and VC5100537582) with black squares whose SFR and $M_{\rm star}$ values fall close to the distribution of the w/o \cii\ Halo objects. 
The existence of these three objects suggest that the selection bias towards luminous objects alone cannot explain the origin of the \cii\ halo. 
\label{fig:ms_test}}
\end{center}
\end{figure}

In Section \ref{sec:results}, we identify individual \cii\ halos surrounding star-forming galaxies at $z=4-6$, 
which is unlikely to be caused by the displacement of the \cii\ line emission from the galaxy. We also identify a similar number of galaxies lacking the \cii\ halo structure. 
Below, the physical origin of the \cii\ halo is explored through a comparison of the physical properties of galaxies with and without the extended \cii\ line morphology. 

\subsection{\cii\ Halo vs. w/o \cii\ Halo Objects}
\label{sec:halo_property} 

\subsubsection{SFR--M$_{\rm star}$ Relation}
\label{sec:discuss1}

In Figure \ref{fig:ms_test}, we examine the SFR--$M_{\rm star}$ relation for the [C {\sc ii}] Halo and w/o [C {\sc ii}] Halo objects separately as classified in Section \ref{sec:cii_halo}. 
For comparison, the entire sample of the 118 ALPINE sources \citep[see][]{schaerer2020} and the Dust \& \cii\ Halo object are also plotted. 

We find that the [C {\sc ii}] Halo and Dust \& \cii\ Halo objects fall in the regime with higher SFR and M$_{\rm star}$ values than the w/o [C {\sc ii}] Halo objects. 
This is consistent with the individual size measurement results that 
show a positive trend of the ratio of $r_{\rm [CII]}$ to $r_{\rm UV}$ as a function of $M_{\rm star}$ (Section \ref{sec:cii-uv_size}) as well as the \cii\ line stacking results with our ALPINE sources, 
where the stacked sample with the high SFR ($>$ 25 $M_{\odot}$ yr$^{-1}$) shows the  \cii\ line structure extended more than the sample with the low ($\leq$ 25 $M_{\odot}$ yr$^{-1}$) SFR \citep{ginolfi2020}.  
These results suggest that the physical origin of the [C {\sc ii}] halo is related to M$_{\rm star}$ and/or SFR through processes such as photoionization and outflows (Section \ref{sec:discuss4}). 

We note that another interpretation for the high SFR and $M_{\rm star}$ trend among the \cii\ Halo objects is a selection bias towards luminous objects whose outskirts of the diffuse emission are easier to to be identified. 
However, there are at least three \cii\ Halo objects (DC630594, DC880016, and VC5100537582; marked with black squares) have the SFR and $M_{\rm star}$ values similar to those of the w/o \cii\ Halo objects. 
In fact, the \cii\ lines from DC880016 and VC5100537582 are detected with SNR at $\sim$ 8 that is comparable to the w/o \cii\ Halo objects of DC539609 and DC733857 (Table \ref{tab:catalog}).   
This indicates that the selection bias alone is hard to explain the difference of the existence of the \cii\ halo. 

\subsubsection{Ly$\alpha$ EW Relation}
\label{sec:discuss2}

\begin{figure}
\begin{center}
\includegraphics[trim=0cm 0cm 0cm 0cm, clip, angle=0,width=0.5\textwidth]{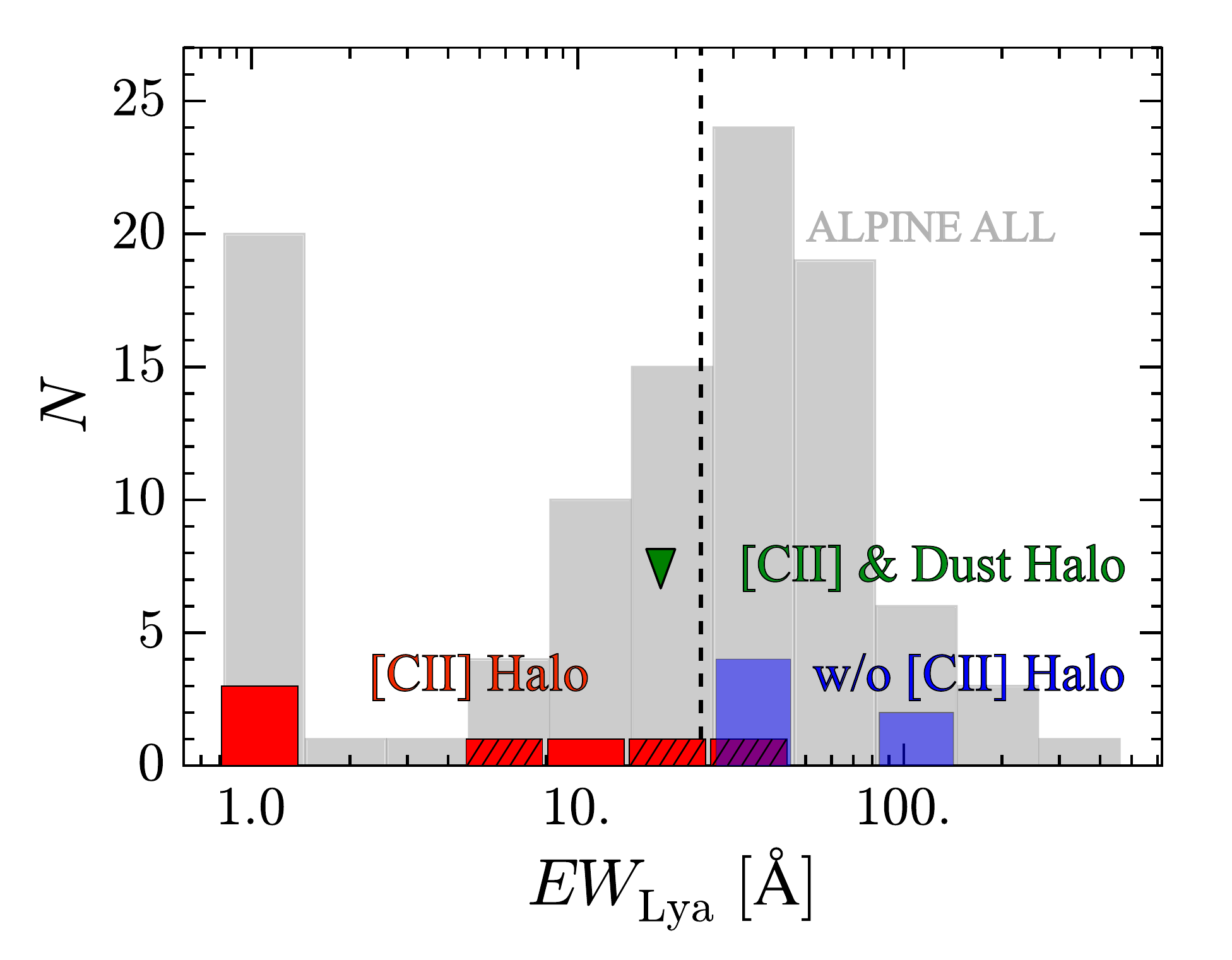}
 \caption[]{
Histogram of the Ly$\alpha$ EW.  
The red and blue histograms indicate the \cii\ Halo and w/o [C {\sc ii}] Halo objects.  
The three \cii\ Halo objects marked with black squares in Figure \ref{fig:ms_test} are shown in the hatched histogram. 
The green reverse triangle denotes the Ly$\alpha$ EW value of the Dust \& \cii\ Halo object. 
The gray histogram shows the entire sample of the 116 ALPINE sources whose Ly$\alpha$ line spectroscopy is available \citep[see][]{cassata2020}. 
The first (leftmost; $\sim1 {\rm \AA}$) bin presents the objects whose Ly$\alpha$ lines are undetected in the optical--NIR spectroscopy. 
The dashed line denotes the median of the entire sample excluding the sources whose Ly$\alpha$ lines are undetected. 
\label{fig:ew_test}}
\end{center}
\end{figure}

In Figure \ref{fig:ew_test}, we investigate the distribution of the Ly$\alpha$ equivalent width (EW) for the [C {\sc ii}] Halo (red histogram), w/o [C {\sc ii}] Halo (blue histogram) objects, and the [C {\sc ii}] \& Dust Halo object (green reverse triangle). 
For comparison, we also show the Ly$\alpha$ EW histogram with the median value (dashed line) 
for the entire sample of the 116 ALPINE sources that are observed spectroscopically for the Ly$\alpha$ emission. 
Note that the first (leftmost; $\sim1 {\rm \AA}$) bin indicates the objects whose Ly$\alpha$ lines are undetected in the optical--NIR spectroscopy.  The details of the Ly$\alpha$ EW measurements are presented in \cite{cassata2020}. 

We find that the \cii\ Halo objects show relatively small Ly$\alpha$ EWs or the Ly$\alpha$ lines undetected, 
while the w/o [C {\sc ii}] Halo objects have relatively large Ly$\alpha$ EWs, compared to the entire sample. 
The Ly$\alpha$ EW decreases for high HI column densities.  
The Ly$\alpha$ EW results thus may indicate that the physical origin of the \cii\ halo is related to the distribution of the neutral hydrogen (Section \ref{sec:discuss4}), 
which is indeed traced by the singly ionized carbon \citep[e.g.,][]{wolfire2003,vallini2015}. 

To remove potential selection biases towards luminous objects, 
we also examine the Ly$\alpha$ EWs for three \cii\ Halo objects of  DC630594, DC880016, and VC510053758 whose SFR and $M_{\rm star}$ values are similar to those of the w/o \cii\ Halo objects (Section \ref{sec:discuss1}). 
In Figure \ref{fig:ew_test}, we show the Ly$\alpha$ EW for these three \cii\ Halo objects in the red hatched histogram. 
We find that the histogram is still likely to have lower Ly$\alpha$ EW values than the w/o \cii\ Halo objects have, 
while one out of three objects fall in the Ly$\alpha$ EW range comparable to the w/o \cii\ Halo objects. 
Although it is hard to draw a definitive conclusion with the small statistics, 
this result suggests that the relatively small Ly$\alpha$ EW trend holds without the potential selection biases, 

In Figure \ref{fig:ew_test}, we also find that the \cii\ \& Dust object has the Ly$\alpha$ EW value of $18 \pm 3\, {\rm \AA}$ which falls relatively lower than the w/o \cii\ Halo objects, but is still close to the median of the entire sample ($\sim 24 \,{\rm \AA}$) even with the existence of the dust spread over the galaxy. This may suggest that the physical origin of the extended structures in both \cii\ and dust emission helps the Ly$\alpha$ line to escape from the system (Section \ref{sec:discuss4}).

\subsubsection{$\Delta v_{\rm Ly\alpha}$ and $\Delta v_{\rm ISM}$ } 
\label{sec:v_off}

\begin{figure}
\begin{center}
\includegraphics[trim=0cm 0cm 0cm 0cm, clip, angle=0,width=0.5\textwidth]{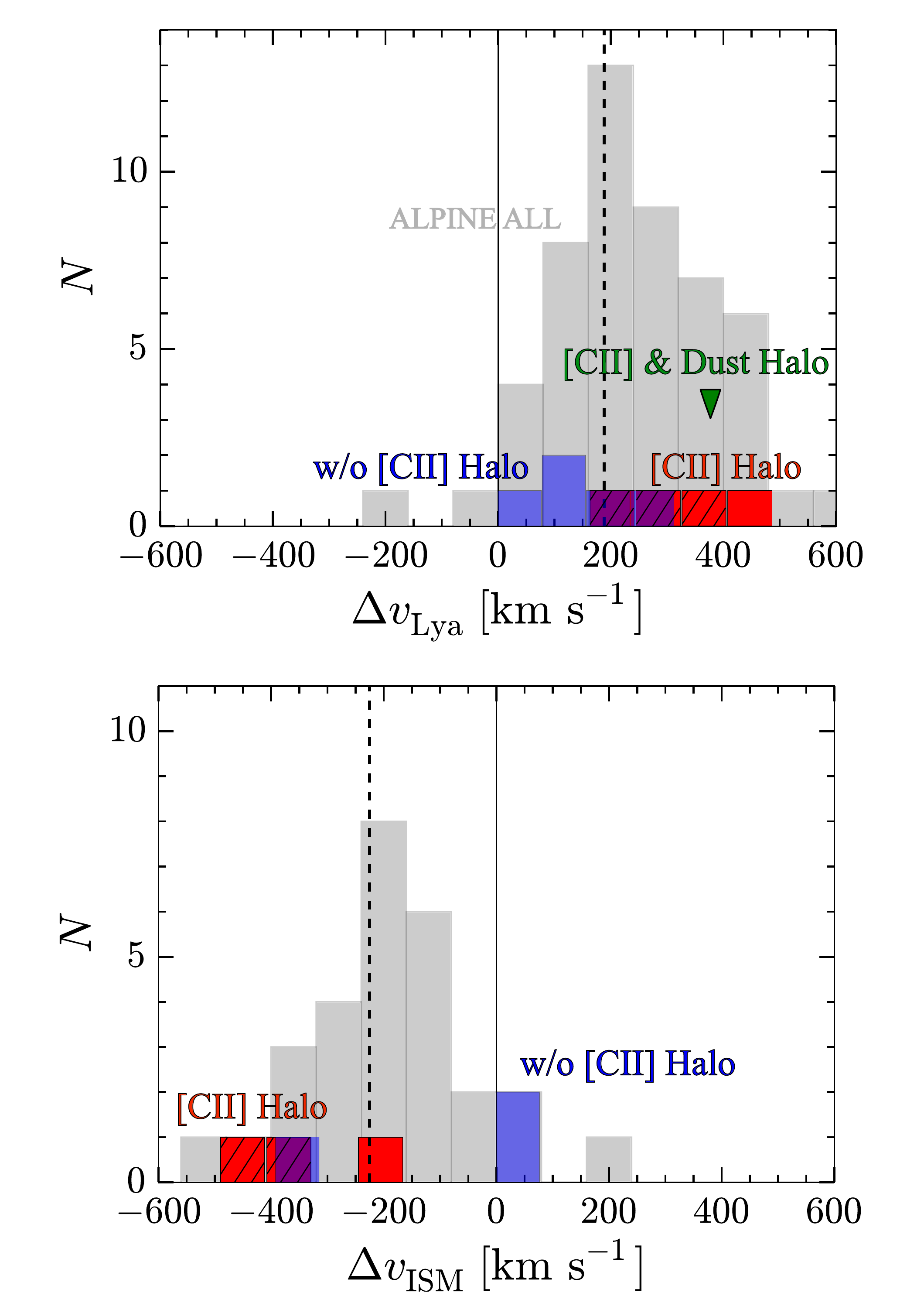}
 \caption[]{
Histograms showing the velocity offsets ($\Delta v$) of the the Ly$\alpha$ line (top) and the rest-frame UV metal absorption (bottom) with respect to the \cii\ line which traces the systemic redshift. 
The red and blue histograms denote the \cii\ Halo and w/o [C {\sc ii}] Halo objects.  
The three \cii\ Halo objects marked with black squares in Figure \ref{fig:ms_test} are presented in the hatched histogram. 
The green reverse triangle in the top panel indicates the  $\Delta v_{\rm Ly\alpha}$ value of the \cii\ \& Dust Halo object. 
The gray histogram shows the entire sample of the 53 and 29 ALPINE sources whose Ly$\alpha$ line and rest-frame UV metal absorption have been identified among the \cii-detected 75 ALPINE sources \citep[see][]{faisst2020, cassata2020}. 
The solid and dashed lines denote the velocity center traced by the \cii\ line and the median value of the entire sample, respectively. 
\label{fig:off_hist}}
\end{center}
\end{figure}

In the top and bottom panels of Figure \ref{fig:off_hist}, we examine the velocity offsets of the Ly$\alpha$ line ($\Delta v_{\rm Ly\alpha}$) and the rest-frame UV metal absorption ($\Delta v_{\rm ISM}$) with respect to the the systemic redshift traced by the \cii\ line, respectively. 
The \cii\ Halo, w/o \cii\ Halo, \cii\ \& Dust Halo object, and the entire sample of the ALPINE sources with its median are shown in the same manner as Figure \ref{fig:ew_test}. 
Note that here we cannot present the sources without the detection of the Ly$\alpha$ line or the rest-frame UV metal absorption. 
The details of the $\Delta v_{\rm Ly\alpha}$ and $\Delta v_{\rm ISM}$ measurements are presented in  \cite{faisst2020} and \cite{cassata2020}. 

The entire sample shows that $\Delta v_{\rm Ly\alpha}$ and  $\Delta v_{\rm ISM}$, in general, take the positive and negative values, corresponding that the Ly$\alpha$ line and the rest-frame UV metal absorption are red- and blue-shifted from the systemic redshift, respectively \citep[see also][]{faisst2020,cassata2020}. 
We find that the \cii\ Halo and \cii\ \& Dust Halo objects typically have larger velocity offsets (more red-shifted Ly$\alpha$ and more blue-shifted rest-frame UV metal absorption) than the w/o [CII] Halo objects. 
The three \cii\ Halo objects of DC630594, DC880016, and VC510053758 whose SFR and $M_{\rm star}$ values are similar to those of the w/o \cii\ Halo objects (red hatched histogram) also show the same trend, albeit with the small statistics. 
The red-shifted Ly$\alpha$ line has been well explained by expanding shell models \citep[e.g.,][]{ahn2004,verhamme2006,verhamme2015}, 
where back-scattered Ly$\alpha$ photons escape from the galaxy owing to the shell experiencing a Doppler shift. 
The blue-shifted rest-frame UV metal absorption is also interpreted as the existence of the outflowing gas between the galaxy and the observer \citep[e.g.,][]{shapley2003,steidel2004,talia2017,sugahara2017,sugahara2019, faisst2020}. 
Although these velocity offset results are obtained from the limited sample whose Ly$\alpha$ line or the rest-frame UV metal absorption has been identified, 
we obtain a consistent picture from these results that ongoing or past episodes of outflows may be linked to the origin of the \cii\ Halo objects (Section \ref{sec:discuss4}). 

\subsubsection{Rotation or Dispersion Dominated?}
\label{sec:cent-kin}

We compare the fractions of \cii\ Halo and w/o \cii\ Halo objects that are either classified as rotator- or dispersion-dominated. 
We find that the \cii\ Halo (w/o \cii\ Halo) objects consist of three (three)  rotator and four (three) dispersion-dominated populations, respectively. 
The nearly equal spread in classification makes it difficult to shed light on the nature of the \cii\ halo.
However, it is worth mentioning that three \cii\ Halo objects of DC630594, DC880016, and VC510053758 whose SFR and $M_{\rm star}$ properties are similar to the w/o \cii\ Halo objects (Section \ref{sec:discuss1}) are all classified as dispersion-dominated. 
This may suggest that some physical mechanisms, especially which increase the velocity dispersion in the system (e.g., past merging events), 
are related to the origin of the \cii\ halo in the ALPINE sources with relatively low SFR and $M_{\rm star}$ values (Section \ref{sec:discuss4}). 

\subsubsection{Kinematics of [CII] Halo}
\label{sec:halo-kin}

\begin{figure*}
\begin{center}
\includegraphics[trim=0cm 0cm 0cm 0cm, clip, angle=0,width=1.0\textwidth]{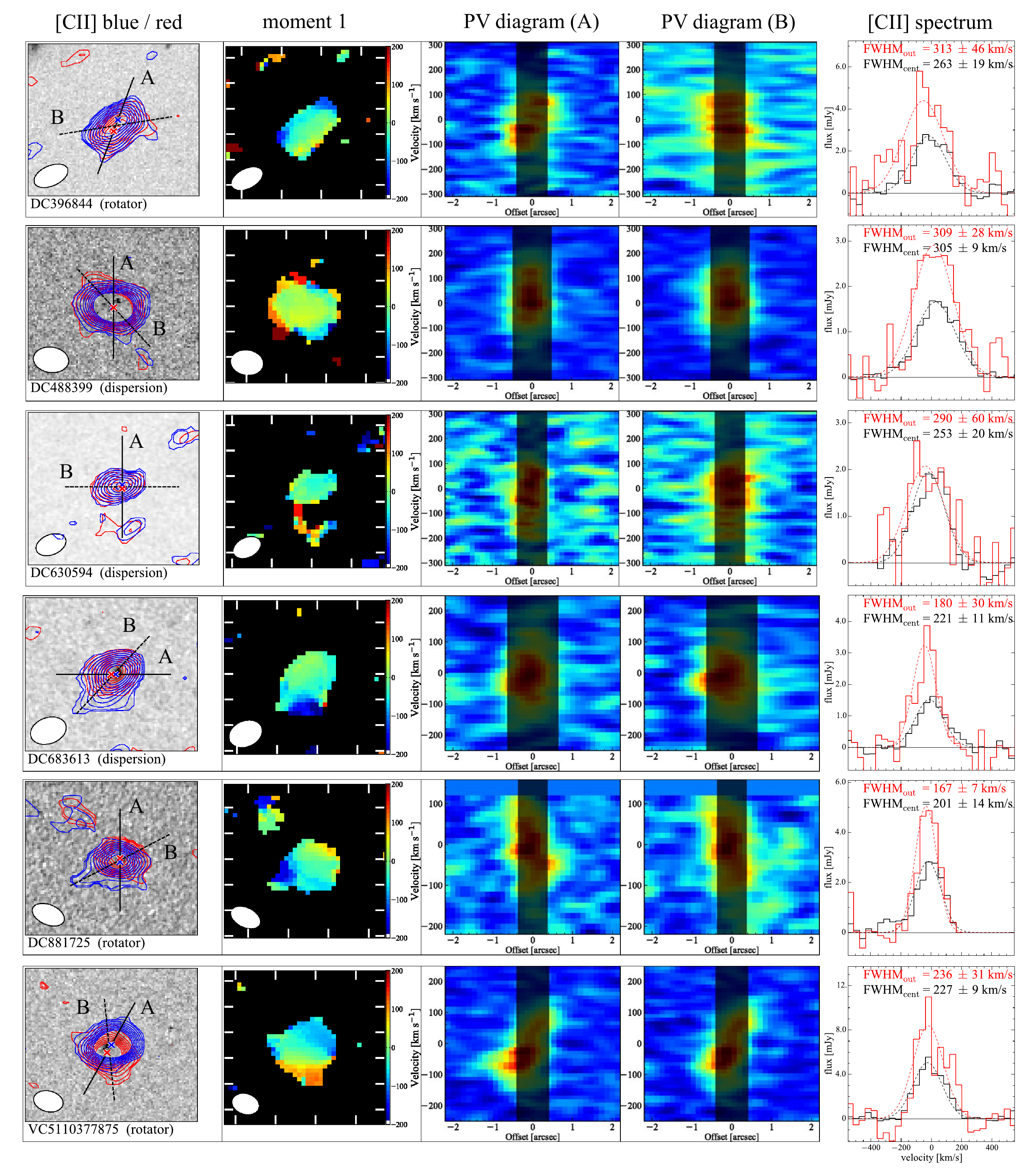}
 \caption[]{
Summary panel of the \cii\ kinematics of the 6 \cii\ Halo objects with SNR $>$ 10. 
{\bf Left to Right}: 
HST rest-frame UV cutout ($6'' \times 6''$) image from ACS F814W or WFC3/IR F160W \citep{koekemoer2007,koekemoer2011} with the blue- (blue contours) and red-shifted (red contours) \cii\ line emission. 
The contours are drawn at $1\sigma$ intervals starting from the 2$\sigma$ level.
The blue and red crosses show the peak pixel positions of the blue- and redshifted \cii\ line maps, respectively. 
The solid line passes the blue and red crosses, while the dashed line is drawn to the major-axis of the \cii\ line morphology.
The central galaxy class (1: rotator, 3: dispersion-dominated, extended, dubbed in ``dispersion'') is presented in the parentheses.; 
Intensity weighted coordinate (moment 1) map.; 
Position-velocity (PV) diagram. 
The two PV diagrams of A and B are obtained along the solid and dashed lines in the left panel, respectively. 
The central area, corresponding to the ALMA beam along to the lines of A or B, is covered by the dark vertical shade for clarity of the \cii\ kinematics at the outer halo area.; 
\cii\ line spectra for the central (black histogram) and outer (red histogram) areas of the galaxy, that are produced from the aperture radius $r_{\rm ap.}$ of  $r_{\rm ap.}\leq0.''5$ ($\simeq$ FWHM/2.0 of ALMA beam) and $0.''5<r_{\rm ap.}\leq1.''6$ ($\simeq$ 10 kpc at $z=5$), respectively. 
We show the best-fit Gaussians (dashed line) and their FWHM values for the central (black) and outer (red) areas in the panel.
\label{fig:halo_kin}}
\end{center}
\end{figure*}

We also investigate the kinematics of the \cii\ line emission, especially in the outer halo areas. 
We first produce the velocity-integrated (moment 0) maps for the blue- and red-shifted \cii\ line emission, 
where the velocity center is defined at the peak frequency in the \cii\ line spectrum. 
We then measure the spatial peak positions of the blue- and red-shifted \cii\ line emission 
and create the position-velocity (PV) diagrams along the lines that pass through these peak positions (A) and the major axis of the \cii\ line emission (B). 
For a complementary approach, we also produce the \cii\ line spectra for the central and outer areas of the objects with the aperture radii $r_{\rm ap.}$ of  $r_{\rm ap.}\leq0.''5$ ($\simeq$ FWHM/2.0 of average ALMA beam) and $0.''5<r_{\rm ap.}\leq1.''6$ ($\simeq$ 10 kpc at $z=5$), respectively. 
In Figure \ref{fig:halo_kin}, 
we summarize the kinematic properties of the \cii\ line emission for the \cii\ Halo and \cii\ \& Dust objects. 
To ensure secure results, we examine only six objects whose \cii\ lines are detected at the $>10\sigma$ level. 
The central area with diameters of $\sim0.''6-1.''2$ (FWHMs of the ALMA beams along to the lines of A and B) is covered by the dark shade in the PV diagrams. 
This helps to visualize the \cii\ kinematics in the outer areas beyond the ALMA beam. 

In the PV diagrams of galaxies classified as dispersion dominated, DC488399, DC630594 and DC683613, we find no clear velocity gradients in the outer area.  
On the other hand, in the galaxies classified as rotators (i.e., DC881725, VC5110377875), we find tentative features of velocity gradient over several beam scales in each case. 
In particular, VC5110377875 has a velocity gradient up to the radius of $\sim2''$ ($\simeq12$ kpc at $z=5.67$) and $\sim -\,100$ km s$^{-1}$ in the PV diagram (A). 
Moreover, the \cii\ line spectrum in the outer area of VC5110377875 shows a double peak profile which is a signature for rotation. 
Since the PV diagram (A) is produced along the line connecting the peak positions of the blue- and red-shifted \cii\ line, 
mostly corresponding to the axis of the galaxy rotation disk, 
the velocity gradient in the outer areas are likely to be associated with the rotation of the central galaxy disk. 
These results imply that 
the kinematics of the \cii\ halo is different between the rotator and dispersion-dominated populations, 
such that for the rotator the extended gas disk is associated and co-rotating with the central galaxy disk as a single large disk.

\begin{figure}
\begin{center}
\includegraphics[trim=0cm 0cm 0cm 0cm, clip, angle=0,width=0.5\textwidth]{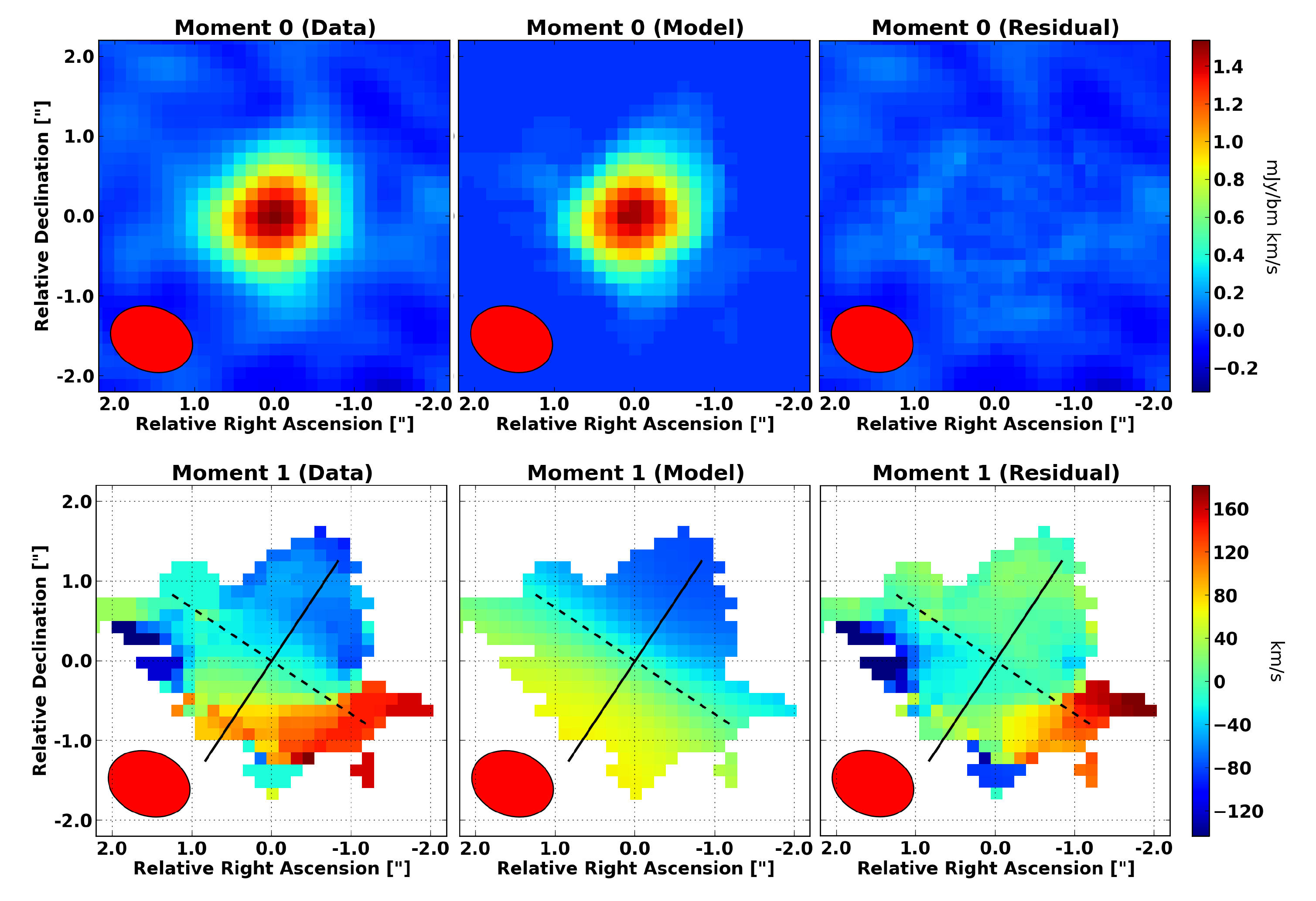}
 \caption[]{
The best-fit 3D modeling results in the moment 0 (top) and moment 1 (bottom) maps for VC5110377875 with $^{\rm 3D}$Barolo \citep{diteodoro2015}. 
The observed, the best-fit model, and the residual maps are presented from left to right. 
The solid and dashed lines are drawn along with the the kinematic major and its perpendicular axes. 
The systematic 3D modeling results for the entire ALPINE sample are presented in Jones et al. in (prep.). 
\label{fig:3d_model}}
\end{center}
\end{figure}

Assuming a single large disk for VC5110377875, we perform 3D modeling with the tilted ring fitting code $^{\rm 3D}$Barolo \citep{diteodoro2015}. 
In Figure \ref{fig:3d_model}, we show the observed, the best-fit model, and the residual maps from left to right for the integrated intensity (moment 0; top) and velocity field (moment 1; bottom). 
The solid and dashed lines are presented along with the kinematic major and its perpendicular axes, respectively.  
We obtain the best-fit effective radius for the single rotating \cii-gas disk of 4.1 $\pm$ 0.5 kpc. 
We note that we use the SEARCH algorithm inside $^{\rm 3D}$Barolo to identify the RMS noise level of the data cube, 
and create a three-dimensional (i.e., RA, Dec, velocity) mask of all signal above 2$\sigma$. 
Since the model is fit only to these masked data, this threshold produces the diffuse residuals (i.e., $\lesssim2\sigma$) at the outskirts in the moment 0 maps.  
Therefore, the best-fit effective radius is the lower limit of size for the rotating \cii-gas disk. 
The rest-frame UV size measurement results show that VC5110377875 has the $r_{\rm e,UV}$ value of $0.93\pm0.3$ kpc (Table \ref{tab:catalog}). 
These results indicate that we identify the rotating \cii-gas disk extended more than the rest-frame UV disk by a factor of $\sim4$ at least. 
Interestingly, we also find symmetric residuals at the outskirts in the moment 1 map mostly along the dashed line. 
This suggests that VC5110377875 features more complex kinematics than could be explained by the single rotating disk. 
If we assume that the rotation of the large \cii-gas disk is aligned to the central galaxy disk, 
the complex kinematics may be caused by the outflow or the existence of another rotating disk perpendicular to the main disk. 
The current data depth is insufficient to conduct the 3D modeling with multiple components for VC5110377875, including the diffuse outskirts and the complex kinematics. 
Deeper observations are essential to investigate the diffuse outskirts and the complex kinematics.

\subsection{Physical Origin of [CII] Halo}
\label{sec:discuss4}
Based on the analysis and the comparison of the physical properties of the \cii\ Halo and w/o [C {\sc ii}] Halo objects investigated in Sections \ref{sec:halo_property},   
here we discuss what could be the origin of the ionized carbon emission in the \cii\ halo. 
One possibility is that the \cii\ halo is always present around star-forming galaxies at $z=4-6$ and the difference in the detection of the \cii\ halo is attributed to a galaxy's position on the SFR--$M_{\rm star}$  relation (Section \ref{sec:discuss1} and Figure \ref{fig:ms_test}).
This induces a selection bias preventing the detection of diffuse \cii\ halo emission around low SFR ($\sim$ faint) ALPINE sources. 
With the fraction of Lyman-alpha emitters decreasing for more luminous ($\sim$ high SFR and $M_{\rm star}$) star-forming galaxies \citep[e.g.,][]{stark2010}, 
this is aligned with the Ly$\alpha$ EW results (Section \ref{sec:discuss2}) showing that the w/o [C {\sc ii}] Halo objects ($\sim$ low SFR) have higher Ly$\alpha$ EWs than the \cii\ Halo objects. 
The high SFR can also contribute to pushing out the gas away from the galaxy more \citep[e.g.,][]{muratov2015}, which is also consistent with the velocity offset results (Section \ref{sec:v_off}). 
However, it is hard to explain the difference of the existence of the \cii\ halo only with this selection bias, 
because we also identify several \cii\ Halo objects (DC630594, DC880016, and VC5100537582; marked with black squares in Figure \ref{fig:ms_test}) with SFR, $M_{\rm star}$, and SNR of the \cii\ line emission similar to those of the w/o \cii\ Halo objects. 
While deeper observations towards the w/o [C {\sc ii}] Halo objects are essential to conclude whether the \cii\ halo homogeneously exists around high-$z$ star-forming galaxies, 
it is worth discussing the possible mechanisms of the \cii\ halo, aside from the selection bias. 
In \cite{fujimoto2019}, the following 5 scenarios are discussed for the origin of the \cii\ halo emission: satellite galaxies, circumgalactic- (CG) scale photodissociation region (PDR), CG-scale H {\sc ii} region, cold streams, and outflow (see Figure 12 and Section 5 in \citealt{fujimoto2019}).
\cite{ginolfi2020b} report the extended \cii\ line structure around a major merging system in a protocluster at $z=4.57$ and suggest another possible scenario of tidal stripping. 
We investigate these 6 ($=5+1$) scenarios based on the physical properties of the \cii\ Halo and w/o [C {\sc ii}] Halo objects analyzed in Sections \ref{sec:halo_property}. 

In the first scenario of satellite galaxies, 
the faint satellite galaxies are observed as an extended structure surrounding the central galaxy. 
Since high SFR $\simeq$ $M_{\rm star}$ objects reside in massive halos potentially with abundant satellite galaxies, 
the trend of the high SFR and $M_{\rm star}$ values in \cii\ Halo objects  (Section \ref{sec:discuss1}) can be ascribed to this scenario. 
Although we use isolated ALPINE galaxies alone in our analysis to reduce the potential contamination of these satellite galaxies, 
the current data depth and resolution might not allow us to detect the faint satellite galaxies. 
However, \cite{fujimoto2019} derive the radial ratio of the \cii\ luminosity to SFR and find in the outer halo area that the ratio becomes much higher ($\gtrsim1$ dex) than the typical ratio of the faint, low-mass galaxies \citep{diaz-santos2014}.
This suggests that the current upper limits of the continuum images in the 
outer halo area already offer us stringent constraints, and the authors rule out the possibility of the satellite galaxies.  
Because the \cii\ Halo galaxies show the radial profiles of the \cii\ line and the continuum emission having the gap in the outer halo area similar to the stacked results in \cite{fujimoto2019}, the satellite galaxy is unlikely to explain the origin of the \cii\ halo emission. 

In the second scenario of CG-scale PDR, the \cii\ Halo objects are surrounded by optically-thick neutral hydrogen, and the CG-scale PDR is formed via the photoionization process. 
The surrounding neutral hydrogen makes the Ly$\alpha$ line hard to escape, which is consistent with the low Ly$\alpha$ EW values in the \cii\ Halo objects (Section \ref{sec:discuss2}). 
In this scenario, the high SFR in the \cii\ Halo objects produces more far-UV (FUV) photons (6 eV $< h\nu<$ 13.6 eV) that penetrate deeper into the surrounding neutral hydrogen and form larger PDR than the w/o [C {\sc ii}] Halo objects of low SFR. 
In these PDRs, the carbon is still singly ionized (ionization potential: 11.3 eV) by the FUV photons \citep[e.g.,][]{hollenbach1999,wolfire2003,vallini2015}, 
which forms the extended \cii\ line structure of the \cii\ halo. 
However, this scenario does not explain the velocity offset results (Section \ref{sec:v_off}), 
indicating that the CG-scale PDR alone is insufficient to the origin of the \cii\ Halo emission. 

In the third scenario of CG-scale H {\sc ii} region, 
the \cii\ Halo objects are producing more ionizing photons or surrounded by optically-thin neutral hydrogen, and the ionizing photons penetrate the ISM and the surrounding CGM deeper (called ``fluorescence" in Ly$\alpha$ halo studies; e.g., \citealt{mas-ribas2016}), where the CG-scale H {\sc ii} region is formed.  
However, more singly ionized carbons (C$^{+}$) are ionized to doubly ionized carbons (C$^{++}$) in the H {\sc ii} region (see also \citealt{ferrara2019}), 
where the \cii\ line emission might become too weak to extend up the CG scale. 
Moreover, the Ly$\alpha$ line easily escapes from the ionized gas, which disagrees with the low Ly$\alpha$ EW values in the \cii\ Halo objects (Section \ref{sec:discuss2}). This scenario is thus unlikely to the origin of the \cii\ halo emission. 

In the fourth scenario of cold streams, 
a dense and cold ($\sim$ 10$^{4}$ K) gas feeds high-$z$ galaxies, as suggested by the cosmological hydrodynamical simulations \citep[e.g.,][]{dekel2009}, and the gravitational energy and shock heating power the \cii\ line emission. 
Since high SFR $\simeq$ $M_{\rm star}$ objects reside in massive halos with potentially high inflow gas rates, 
the trend of the high SFR and $M_{\rm star}$ values in \cii\ Halo objects (Section \ref{sec:discuss1}) can be consistent with this scenario. 
The high inflow gas rate may also contribute to forming the optically-thick neutral hydrogen surrounding the central galaxy and cause the low Ly$\alpha$ EW values in the \cii\ Halo objects (Section \ref{sec:discuss2}). 
However, given the generally low metallicity of the inflowing gas transferred from the intergalactic medium (e.g., $\sim10^{-3}Z_{\odot}$ at $z\sim4$--6; see \citealt{pallottini2014b}), 
this is not supposed to contribute much to the extended \cii\ halo emission because of the low emissivity of the \cii\ line in such low-metallicity gas \citep[e.g.,][]{vallini2015}.

In the fifth scenario of outflow, the ionized carbon powered by the AGN or star formation (SF) feedback form the \cii\ halo, where the associated process of shock heating \citep[e.g.,][]{appleton2013} may also contribute to radiating the extended \cii\ halo emission.
\cite{ginolfi2020} report a significant detection of a broad feature in the wings of the stacked \cii\ line spectrum, 
which is a probe for on-going outflow activities \citep[e.g.,][]{maiolino2012, cicone2015,gallerani2018, bischetti2019, stanley2019}, 
that is more prominent when stacking ALPINE sources at high SFRs. 
From an independent approach, 
\cite{faisst2020} show that the stacked rest-frame UV metal absorption is more significantly blue-shifted in the ALPINE sources with high specific SFR ($\equiv {\rm SFR}$/$M_{\rm star}$; sSFR) than in the sources with low sSFR. 
These results suggest the existence of SF-driven outflows in these galaxies, 
which is consistent with our results that the \cii\ Halo objects are generally seen in galaxies with high values of SFR and $M_{\rm star}$ (Section \ref{sec:discuss1}) and that the \cii\ Halo objects are likely to have more red-shifted Ly$\alpha$ and blue-shifted rest-frame UV metal absorption than the w/o \cii\ Halo objects  (Section \ref{sec:v_off}).
The fraction of Ly$\alpha$ emitters decreases in these high SFR ($\simeq$ luminous) star-forming galaxies \citep[e.g.,][]{stark2010}, which is consistent with the low Ly$\alpha$ EW values in the \cii\ Halo objects (Section \ref{sec:discuss2}). 
The SF-driven outflow may also explain the high fraction of the dispersion-dominated population for the \cii\ Halo objects with relative low SFR and $M_{\rm star}$ values (Section \ref{sec:cent-kin}). 
The past merging events make the system dispersion-dominated and induce the intense star-forming activity, 
where the SF-driven outflow pushes out the ionized carbon outside the galaxy. 
As the amount of gas declines, the SFR value decreases as well. We then witness the extended \cii\ emission as the remnant of the past SF-driven outflow possibly around poststarburst galaxies. 
The 3D modeling results for VC5110377875 show the symmetric residuals which can also be explained by the outflow (Section \ref{sec:halo-kin}).  
In addition to these agreements with the observational results,
the recent theoretical results also support the SF-driven outflow. 
\cite{pizzati2020} model the \cii\ line emission produced from the supernova (SN) driven cooling outflow with two key parameters of the outflow mass loading factor $\eta$ and circular velocity of the parent galaxy dark matter halo $v_{\rm circ}$ that depends on the gravitational potential of the system. 
The authors find that the 10-kpc scale \cii\ halo is well reproduced with the best-fit parameters of $\eta=3.20 \pm 0.10$ and $v_{\rm circ} = 170\pm10$ km s$^{-1}$ around a galaxy with the dynamical mass of $\simeq10^{11}\,M_{\odot}$. 
With cosmological hydrodynamic simulations and radiative transfer calculations, 
\cite{arata2020} derive the time evolution of $r_{\rm e, [CII]}$ and find that $r_{\rm e, [CII]}$ becomes extended in the SF-driven outflow phase.  

In the last scenario of tidal stripping, 
strong dynamical interactions occur between merging galaxies \citep[e.g.,][]{ginolfi2020b}, 
where the shock heating powers the \cii\ line emission \citep[e.g.,][]{appleton2013} in the outer halo area. 
Since we focus on the isolated ALPINE galaxies (i.e., not classified as mergers) and confirm in the discussion of the first scenario that the ratio of the \cii\ line and continuum in the outer halo area rules out the existence of faint, low-mass satellite galaxies, the tidal stripping is unlikely the major contributor of the \cii\ halo emission around the isolated galaxy. 

Based on these discussions, 
the SF-driven outflow is the most likely to be the origin of the \cii\ halo, which can explain all physical properties of the \cii\ Halo and w/o \cii\ Halo objects analyzed in Section \ref{sec:halo_property}. 
There are two types of outflow, hot-mode and cold-mode \citep[e.g.,][]{murray2011,hopkins2014,muratov2015,heckman2017}. 
The hot-mode outflow is defined as the outflow of ionized hydrogen ($\simeq10^{6-7}$ K) gas heated by supernova (SN) explosions, massive star/AGN radiation, while the cold-mode outflow consists of the cold neutral hydrogen gas ($\simeq10^{2-4}$ K) pushed by the radiative and kinetic pressures exerted by SNe, massive stars, and AGNs. 
The SF-driven outflow, seen in ALPINE sources, may be dominated by the cold-mode outflow given the requirement for optically-thick neutral hydrogen to explain the Ly$\alpha$ EWs discussed above.
We note that this is a simplified picture. A realistic picture might have a transition from hot to cold through a catastrophic cooling in the outflow process \citep{li2019,pizzati2020}.
The study for these multi-phase (cold and hot) gas structures will be useful to understand which is the dominant mode and constrain the feedback mechanisms in the high-$z$ star-forming galaxies. 

As discussed in previous studies \citep{fujimoto2019,ginolfi2020}, the outflow activities are essential in the most of the scenarios to enrich the CGM with carbon around the isolated star-forming galaxies in the early universe at $z=4-6$.  
The tidal stripping might also contribute to the CGM metal enrichment even around the isolated galaxy, if the galaxy experienced past merging events. 
In these outflow or tidal stripping, widespread dust can also exist, whereas the majority of the ALPINE sources do not show extended structure in the rest-frame FIR continuum. 
This is probably because such dust quickly becomes cold in the outer galaxy and its intrinsic morphology is then a challenge to be observed, especially at high-redshift due to the warm CMB \citep[e.g.,][]{dacunha2013,vallini2015,zhang2016,pallottini2017b,lagache2018}.  
Interestingly, we do identify one \cii\ \& Dust Halo object whose morphology is extended in both the \cii\ line and the rest-frame FIR continuum, but compact in the rest-frame UV continuum (Section \ref{sec:cii_halo}). 
This may indicate that some mechanisms, such as the hot-mode outflow, keep the dust hot even in the outer galaxy, 
and its extended morphology is observed in the \cii\ \& Dust Halo object.
Since the escape fraction of the Ly$\alpha$ line increases (decreases) in the ionized gas (in the gas with dust), the fact that the \cii\ \& Dust Halo object has the Ly$\alpha$ EW close to the typical value among the ALPINE sources (Section \ref{sec:discuss2}), rather than much lower values with the existence of the dust spread over the galaxy, may be contributed by the hot-mode outflow. 
Note, in principle, that faint dusty satellite galaxies can also explain the extended morphology of the rest-frame FIR continuum. 
However, assuming a mass--metallicity relation \citep[e.g.,][]{mannucci2010}, 
the faint ($\simeq$ low mass) satellite galaxies are likely to be less dusty \citep[c.f.,][]{faisst2017}. 
If faint dusty galaxies are identified in future deep observations, they will provide 
important insight to understand the metal enrichment process in early galaxy formation and evolution.

\section{Summary}
\label{sec:summary}
In this paper, we study the detailed morphology of {\sc [Cii]} line emission for 46 main-sequence star-forming galaxies at $z=4-6$, 
from the large ALMA project ALPINE, whose \cii\ lines are individually detected above $5\sigma$ level.  
In conjunction with the HST images, 
we examine the radial surface brightness profiles of the {\sc [Cii]} line, rest-frame FIR, and UV continuum emission. 
We then discuss the possible physical origin of the extended {\sc [Cii]} line emission. 
The major findings of this paper are summarized below: 

\begin{enumerate}
\item We perform size measurements of the \cii\ line in the $uv$-visibility plane, assuming an exponential light profile, and find that the median effective radius is 2.1 $\pm$ 0.16 kpc for the ALPINE sources with $M_{\rm star}$  $\sim10^{9}-10^{10.5}\,M_{\odot}$. Using the same exponential profiles for the HST data in the rest-frame UV wavelength, we find that the \cii\ line size is almost always larger than the rest-frame UV continuum, and that the ratio of \cii\ over the rest-frame UV effective radii as a function of $M_{\rm star}$ shows an increasing trend in the range from $\sim2$ to 3. 
\item  A number of the isolated ALPINE sources show an extended \cii\ halo structure extending up to a radius of $\sim$ 10\,kpc or more, while being compact in continuum emission. We evaluate the significance level of the extended emission and identify $\sim 30 \%$ of the isolated ALPINE sources whose \cii\ line emission extended over the ALMA beam is detected above the $4\sigma$ level, but the rest-frame UV and FIR continuum are below the $3\sigma$ level in the same aperture (\cii\ Halo object). We also identify galaxies without a \cii\ halo structure at a similar fraction as compared to the isolated ALPINE sources whose \cii\ line emission extended over the ALMA beam is below the $3\sigma$ level (w/o \cii\ Halo object).   
\item We examine the spatial offsets among the \cii\ line, the rest-frame FIR, and the rest-frame UV continuum emission peaks for the isolated ALPINE sources. The median (standard deviation) of the offsets are estimated to be $0\farcs15$  ($0\farcs09$), $0\farcs25$  ($0\farcs08$), and $0\farcs15$  ($0\farcs12$), between \cii\ --UV, FIR--UV, and \cii\ --FIR, respectively, corresponding to the $\sim$1 kpc scale at $z=4-6$. These offsets are comparable to the uncertainties in the HST astrometry and in the positional accuracy of the ALPINE sources thus raising a bit of caution that these emitting regions might not always be placed significantly away from each other as being reported in the literature. 
\item We compare the physical properties of the \cii\ Halo and w/o [C {\sc ii}] Halo objects. We find that the \cii\ Halo objects generally have higher SFR and $M_{\rm star}$, more blue-shifted (red-shifted) rest-frame UV metal absorption (Ly$\alpha$ line), but lower Ly$\alpha$ EW than the w/o \cii\ Halo objects. We also find no clear difference between \cii\ Halo and w/o \cii\ Halo objects in the kinematics of the central galaxy, rotation or dispersion dominated.  Among \cii\ Halo objects, the \cii\ kinematics in the outer halo areas may be associated and co-rotating with the central galaxy disk, when the central galaxy is a rotator. We tentatively identify rotation features up to $\rm 10$ kpc in VC5110377875, where the 3D model fit results show that the rotating \cii-gas disk extends over at least 4 times larger than the rest-frame UV emitting region. 
\item From the comparisons of the physical properties of \cii\ Halo and w/o \cii\ Halo objects, one possibility is that the \cii\ halo exists around all star-forming galaxies at $z=4-6$ with the differences in the  SFR -- $M_{\rm star}$ plane and the Ly$\alpha$ EW distribution caused by a selection bias towards the luminous objects. However, we also identify at least three \cii\ Halo objects whose \cii\ line luminosity, SFR, and $M_{\rm star}$ values are similar to the w/o \cii\ Halo objects, suggesting that the difference in the physical properties cannot be explained by the selection bias alone. We discuss the following six scenarios for the physical origin of the \cii\ halo: satellite galaxies, CG-scale PDR, CG-scale H {\sc ii} region, cold stream, outflow, and tidal stripping. We find that the star-formation driven outflow can explain all trends in the physical properties of \cii\ Halo and w/o \cii\ Halo objects and thus is the most likely origin of the \cii\ halo, aside from the selection bias.  

\end{enumerate}

We thank Tanio D\'iaz-Santos, Andrea Ferrara, Simona Gallerani, Akio Inoue, Hiroshi Nagai, Kentaro Nagamine, Kimihiko Nakajima, Masami Ouchi, Takatoshi Shibuya, and Francesco Valentino  (in alphabetical order) for useful comments and suggestions. 
This paper makes use of the ALMA data: ADS/JAO. ALMA \#2017.1.00428.L. 
ALMA is a partnership of the ESO (representing its member states), 
NSF (USA) and NINS (Japan), together with NRC (Canada), MOST and ASIAA (Taiwan), and KASI (Republic of Korea), 
in cooperation with the Republic of Chile. 
The Joint ALMA Observatory is operated by the ESO, AUI/NRAO, and NAOJ. 
This work is based on observations and archival data made with the Spitzer Space Telescope, which is operated by the Jet Propulsion
Laboratory, California Institute of Technology, under a contract with NASA along with archival data from the NASA/ESA Hubble
Space Telescope. This research made also use of the NASA/IPAC Infrared Science Archive (IRSA), 
which is operated by the Jet Propulsion Laboratory, California Institute of Technology, under contract with the National Aeronautics and Space Administration. 
In parts based on data products from observations made with ESO Telescopes at the La Silla Paranal Observatory under ESO programme ID 179.A-2005 
and on data products produced by TERAPIX and the Cambridge Astronomy Survey Unit on behalf of the UltraVISTA consortium. 
Based on data obtained with the European Southern Observatory Very Large Telescope, Paranal, Chile, under Large Program 185.A-0791, 
and made available by the VUDS team at the CESAM data center, Laboratoire d'Astrophysique de Marseille, France. 
This work is based on observations taken by the 3D-HST Treasury Program (GO 12177 and 12328) with the NASA/ESA HST, 
which is operated by the Association of Universities for Research in Astronomy, Inc., under NASA contract NAS5-26555. 
Furthermore, this work is based on data from the W.M. Keck Observatory and the Canada-France-Hawaii Telescope, 
as well as collected at the Subaru Telescope and retrieved from the HSC data archive system, 
which is operated by the Subaru Telescope and Astronomy Data Center at the National Astronomical Observatory of Japan. 
Finally, we would also like to recognize the contributions from all of the members of the COSMOS Team who
helped in obtaining and reducing the large amount of multiwavelength data that are now publicly available through IRSA
at http://irsa.ipac.caltech.edu/Missions/cosmos.html.
This study is supported by the NAOJ ALMA Scientific Research Grant Number 2016-01A.
S.F. acknowledge support from the European Research Council (ERC) Consolidator Grant funding scheme (project ConTExt, grant No. 648179). 
The Cosmic Dawn Center is funded by the Danish National Research Foundation under grant No. 140.
J.D.S. was supported by the JSPS KAKENHI Grant Number JP18H04346, and the World Premier International Research Center Initiative (WPI Initiative), MEXT, Japan.
E.I.\ acknowledges partial support from FONDECYT through grant N$^\circ$\,1171710. 
A.C., C.G., F.L., F.P., M.T.  acknowledge the support from grant PRINMIUR 2017-20173ML3WW\_001. 
L.V. acknowledges funding from the European Union’s Horizon 2020 research and innovation program under the Marie Sklodowska-Curie Grant agreement No. 746119. 
G.C.J. and R.M. acknowledge ERC Advanced Grant 695671 ``QUENCH'' and support by the Science and Technology Facilities Council (STFC).

\bibliographystyle{apj}
\bibliography{apj-jour,reference}

\clearpage

\appendix
\section{A. Additional ALPINE Sources for the rest-frame UV size measurement} 
\label{sec:appendix_uv}

\begin{table*}[h]
{\scriptsize
\caption{Our ALPINE Source Catalog (3.5 $\leq$ SNR $<$ 5)}
\label{tab:catalog2}
\begin{center}
\begin{tabular}{lccccccc}
\hline
\hline
Name  & $z_{\rm [CII]}$ & SNR & $r_{\rm e, f814w}$ & flag$_{\rm f814w}$ & $r_{\rm e, f160w}$ & flag$_{\rm f160w}$ & morph. class     \\
      &                  &     &  (kpc)              &                     &       (kpc)         &                     &                  \\
(1)   &    (2)           & (3) &            (4)      &     (5)             &    	(6)           &           (7)       &         (8)      \\
\hline
CG12 & 4.431 & 4.4 & 0.62 $\pm$ 0.09 & 0 & 1.02 $\pm$ 0.04 & 0 & 5 \\
CG14 & 5.5527 & 4.6 & 0.51 $\pm$ 0.07 & 1 & 0.92 $\pm$ 0.03 & 0 & 4 \\
CG21 & 5.5716 & 4.2 & 0.98 $\pm$ 0.32 & 0 & 1.2 $\pm$ 0.14 & 0 & 4 \\
CG38 & 5.5721 & 4.7 & 1.08 $\pm$ 0.33 & 0 & 1.99 $\pm$ 0.1 & 1 & 2 \\
CG42 & 5.5252 & 3.7 & 0.64 $\pm$ 0.23 & 0 & 1.4 $\pm$ 0.16 & 1 & 5 \\
CG47 & 5.5745 & 4.0 & 1.01 $\pm$ 0.25 & 0 & 1.57 $\pm$ 0.09 & 1 & 4 \\
CG75 & 5.5666 & 4.8 & 0.87 $\pm$ 0.36 & 0 & 1.76 $\pm$ 0.22 & 1 & 4 \\
DC274035 & 4.4791 & 4.4 & 0.91 $\pm$ 0.23 & 0 & No data & -- & 4 \\
DC378903 & 5.4297 & 4.6 & 0.52 $\pm$ 0.43 & 1 & 0.63 $\pm$ 0.41 & 0 & 2 \\
DC400160 & 4.5404 & 4.5 & 1.01 $\pm$ 0.19 & 0 & No data & -- & 4 \\
DC430951 & 5.6881 & 4.1 & 0.51 $\pm$ 0.32 & 0 & No data & -- & 5 \\
DC510660 & 4.548 & 4.0 & 0.95 $\pm$ 0.25 & 1 & No data & -- & 5 \\
DC628063 & 4.5327 & 3.9 & 0.75 $\pm$ 0.3 & 1 & 0.93 $\pm$ 0.48 & 1 & 5 \\
DC665509 & 4.5256 & 4.8 & 1.07 $\pm$ 0.24 & 1 & 0.79 $\pm$ 0.28 & 1 & 2 \\
DC665626 & 4.5773 & 4.4 & 0.0 $\pm$ 654.72 & 1 & No data & -- & 5 \\
DC680104 & 4.5295 & 4.2 & 0.84 $\pm$ 0.27 & 0 & No data & -- & 5 \\
DC803480 & 4.5417 & 3.8 & 0.59 $\pm$ 0.15 & 0 & No data & -- & 4 \\
DC814483 & 4.581 & 4.7 & 1.14 $\pm$ 0.28 & 1 & No data & -- & 2 \\
DC843045 & 5.8473 & 4.1 & 0.75 $\pm$ 0.45 & 0 & 1.08 $\pm$ 0.49 & 1 & 2 \\
DC859732 & 4.5318 & 4.3 & 0.0 $\pm$ 657.75 & 1 & No data & -- & 2 \\
vc5101209780 & 4.5701 & 4.3 & 1.0 $\pm$ 0.24 & 0 & No data & -- & 2 \\
vc5101210235 & 4.5761 & 4.3 & 1.01 $\pm$ 0.19 & 0 & No data & -- & 1 \\
vc5101288969 & 5.7209 & 4.2 & 0.08 $\pm$ 1.7 & 1 & 0.83 $\pm$ 0.39 & 0 & 5 \\
vc510605533 & 4.5019 & 4.9 & 0.65 $\pm$ 0.19 & 0 & 0.28 $\pm$ 4.71 & 1 & 5 \\
DC357722 & 5.6838 & 3.6 & 0.37 $\pm$ 2.32 & 1 & 0.72 $\pm$ 2.64 & 1 & 4 \\
DC722679 & 5.7168 & 4.0 & 0.69 $\pm$ 0.25 & 0 & No data & -- & 5 \\
DC742174 & 5.636 & 4.8 & 0.2 $\pm$ 2.01 & 1 & 0.61 $\pm$ 0.44 & 1 & 5 \\
DC845652 & 5.3071 & 4.9 & 0.4 $\pm$ 0.1 & 0 & 0.83 $\pm$ 0.07 & 1 & 5 \\
\hline
\end{tabular}
\end{center}
\tablecomments{
\footnotesize{
(1) ALPINE source name. We refer to CANDELS$\_$GOODS, DEIMOS$\_$COSMOS, VUDS$\_$COSMOS, and VUDS$\_$ECDFS in \cite{lefevre2019} as CG, DC, VC, and VE, respectively.  
We list 29 ALPINE sources with SNR of the \cii\ line in the range from 3.5 to 5.0 that are used for the rest-frame UV size measurement in this paper. 
The entire sample of 118 ALPINE sources with the coordinate is presented in \citealt{lefevre2019}. 
(2) Spectroscopic redshift estimated from the \cii\ 158$\mu$m line.  
(3) The peak SNR in the velocity-integrated map. 
(4) Circularized effective radius of the rest-frame UV emission in the HST/F814W map measured with {\sc galfit} (see text). 
(5) Flag for the reliability of the {\sc galfit} fitting for the F814W map. 
The same flag definition as (5). 
(6) Circularized effective radius of the rest-frame UV emission in the HST/F160W map measured with {\sc galfit} (see text). 
(7) Flag for the reliability of the {\sc galfit} fitting for the F160W map.  
The same flag definition as (5). 
(8) Galaxy type based on the morphology+kinematic classification (1: rotator, 2: pair-merger, 3: dispersion-dominated, extended, 4: dispersion-dominated, compact, 5: weak; see \citealt{lefevre2019}). 
}}
}
\end{table*}

\clearpage

\section{B. Spatial Distributions and Radial Profiles of the \cii\ line, the rest-frame FIR, and UV continuum for Isolated ALPINE Sources}
\label{sec:appendix_postage}

\begin{figure*}[h]
\begin{center}
\includegraphics[trim=0cm 0cm 0cm 0cm, clip, angle=0,width=1.\textwidth]{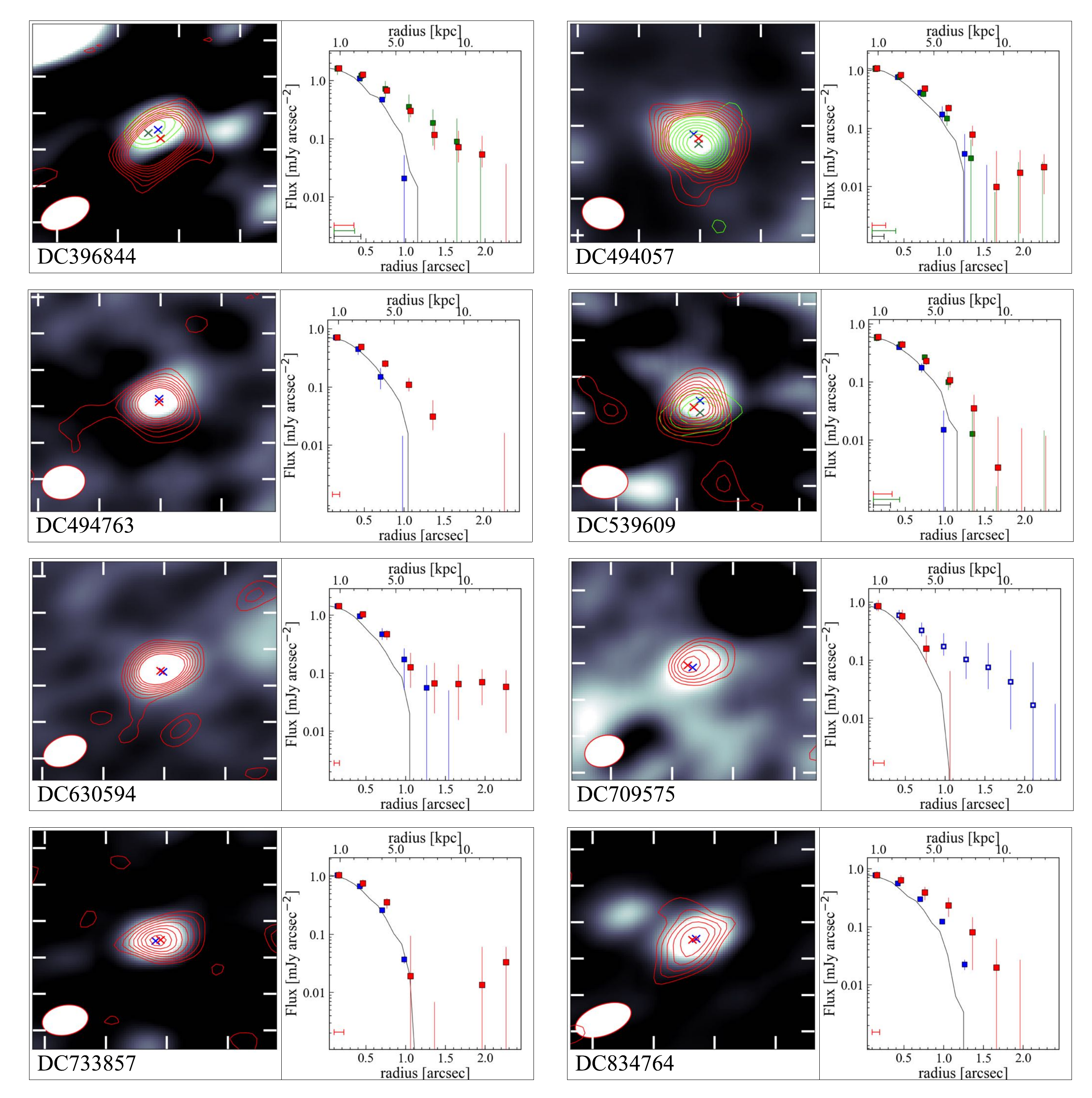}
 \caption[]{
 Spatial distributions of the \cii\ line, the rest-frame FIR and UV continuum for the ALPINE sources whose \cii\ lines are detected above $5\sigma$ level and that are not classified as mergers (Section \ref{sec:product}). 
 The color and symbols are in the same assignment as Figure \ref{fig:postage_radial}. 
\label{fig:postage_ap}}
\end{center}
\end{figure*}

\begin{figure*}
\begin{center}
\includegraphics[trim=0cm 0cm 0cm 0cm, angle=0,width=1.\textwidth]{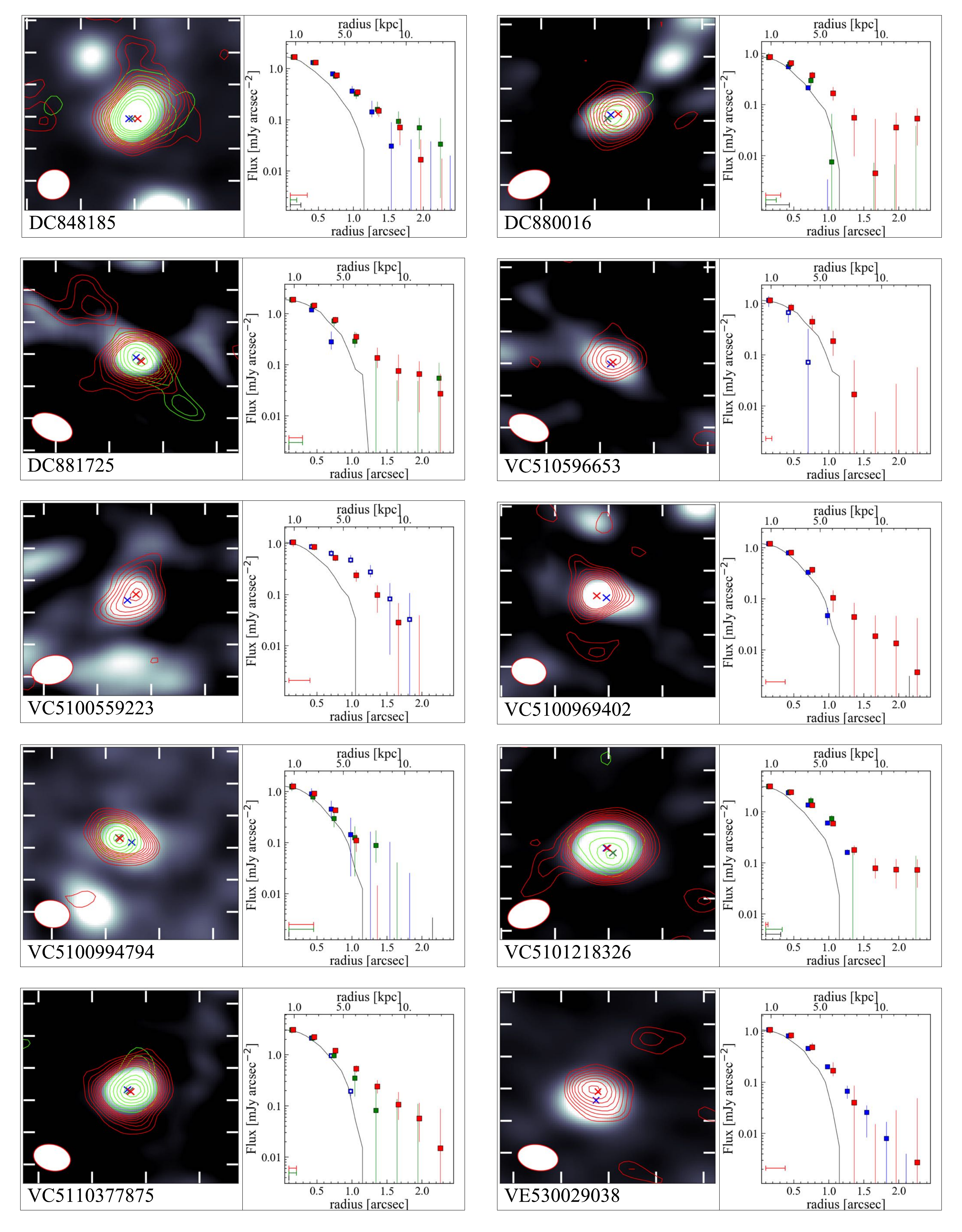}
{\bf Figure \ref{fig:postage_ap}.} (continued)
\end{center}
\end{figure*}

\end{document}